\documentclass[12pt]{article}
\usepackage[margin=1in]{geometry}

\usepackage{amsmath,amsthm,amssymb,eurosym}
\usepackage{enumerate}
\usepackage{graphicx}
\usepackage{fancyhdr}
\usepackage{lastpage}
\usepackage{titlesec}
\usepackage[OT2,T1]{fontenc} 
\usepackage[russian,american]{babel} 
\usepackage{float}
\usepackage{bbm}
\usepackage[usenames,dvipsnames]{xcolor}
\usepackage{color}
\usepackage{soul} %for highlighting
\usepackage{wrapfig}
\usepackage{verbatim}
\usepackage{algorithm}
\usepackage{algorithmic}
\usepackage{framed} % for shaded boxes
\usepackage[utf8]{inputenc}
\usepackage[T1]{fontenc}
\usepackage{csquotes} % for block quotes
\usepackage{url}
\renewcommand{\P}{\mathbf{P}} 	% probability
\newcommand{\E}{\mathbf{E}} 	% expectation
\newcommand{\var}{\operatorname{Var}}   % variance
\newcommand{\Var}{\operatorname{Var}}   % variance
\newcommand{\cov}{\operatorname{Cov}} 	% covariance
\newcommand{\cY}{\mathcal{Y}}

\newcommand{\cS}{\mathcal{S}}

\newcommand{\one}{\mathbbm{1}}  % indicator variable

\newcommand{\ep}{\varepsilon}

\newcommand{\on}[1]{\operatorname{#1}}

\newcommand{\wto}{\Rightarrow}
\newcommand{\sumin}{\sum_{i=1}^N} 	%sum from i=1 to n
\theoremstyle{plain}

\newtheorem{proposition}{Proposition}
\newtheorem{assumption}{Assumption}
\theoremstyle{definition}

\newtheorem{remark}{Remark}

\usepackage{booktabs}
\usepackage{multirow}
\usepackage[colorlinks=false,hidelinks]{hyperref}
\usepackage{chngcntr}

\providecommand{\keywords}[1]{\textit{Keywords: } #1}

\newcommand{\sumi}{\sum_{i=1}^N}

\newcommand{\avgin}{\frac{1}{N} \sumi}

\newcommand{\taufp}{\tau_{\text{fp}}}
\newcommand{\tausp}{\tau_{\text{sp}}}

\newcommand{\neff}{n_{\text{eff}}}
\newcommand{\neffpop}{n^*_{\text{eff}}}

\newcommand{\DM}{\hat \tau_{\text{DM}}}

\newcommand{\paluck}[1]{#1}

\usepackage{setspace}
\usepackage{titling}
\usepackage{natbib}
\setcitestyle{authoryear,aysep={}}%,open={(},close={)}}

\title{Evaluating stochastic seeding strategies in networks\thanks{Authors are listed in alphabetical order. We thank Jing Cai, Elizabeth Paluck, Hana Shepherd, and Peter Aronow for advice in working with data from their field experiments. We thank Hao Yin for suggesting the dynamic programming solution discussed in Appendix~\ref{app:probs}. This work benefited from comments by Sinan Aral, John Hauser, Maurits Kaptein, M. Amin Rahimian, Duncan Simester, Juanjuan Zhang, Yunhao Zhang, two anonymous referees, and seminar participants at Carnegie Mellon University, Columbia University, Duke University, Harvard University, New York University, University of California Los Angeles, University of Massachusetts Amherst, University of North Carolina, University of Southern California, University of Toronto, and Yale University. This work was supported in part by NSF grant IIS-1657104.}
}
\date{\small Working paper. This version: \today. \vspace{-20pt}}
\author{
Alex Chin\thanks{Department of Statistics, Stanford University.}
\and
Dean Eckles\thanks{Sloan School of Management, Massachusetts Institute of Technology.} \thanks{Institute for Data, Systems \& Society, Massachusetts Institute of Technology.}
\thanks{To whom correspondence should be addressed: eckles@mit.edu.}
\and
Johan Ugander\thanks{Management Science \& Engineering Department, Stanford University.} \thanks{Institute for Computational \& Mathematical Engineering, Stanford University.}}

\begin{document}
\maketitle

\begin{abstract}
When trying to maximize the adoption of a behavior in a population connected by a social network, it is common to strategize about where in the network to seed the behavior, often with an element of randomness. Selecting seeds uniformly at random is a basic but compelling strategy in that it distributes seeds broadly throughout the network. A more sophisticated stochastic strategy, one-hop targeting, is to select random network neighbors of random individuals; this exploits a version of the friendship paradox, whereby the friend of a random individual is expected to have more friends than a random individual, with the hope that seeding a behavior at more connected individuals leads to more adoption. Many seeding strategies have been proposed, but empirical evaluations have demanded large field experiments designed specifically for this purpose and have yielded relatively imprecise comparisons of strategies. Here we show how stochastic seeding strategies can be evaluated more efficiently in such experiments, how they can be evaluated ``off-policy'' using existing data arising from experiments designed for other purposes, and how to design more efficient experiments. In particular, we consider contrasts between stochastic seeding strategies and analyze nonparametric estimators adapted from policy evaluation and importance sampling. We use simulations on real networks to show that the proposed estimators and designs can increase precision while yielding valid inference. We then apply our proposed estimators to two field experiments, one that assigned households to an intensive marketing intervention and one that assigned students to an anti-bullying intervention.  
\end{abstract}

\pagebreak
%\onehalfspacing

%\tableofcontents

\section{Introduction}
Interventions are often targeted to individuals based on their observed characteristics~\citep{manski2004statistical,simester2020efficiently}.
Given the large theoretical and empirical literature showing the prevalence of peer effects in adoption processes, it is reasonable to assume that targeting strategies that incorporate network characteristics may be particularly successful. For example, a simple strategy might be to target an intervention at a small number of ``seed individuals'' who are well-connected and centrally located in the social network of the target population. Such a strategy aims to go beyond direct targeting and identify individuals who are expected to have an outsized impact on total adoption \citep{krackhardt1996structural,kempe2003maximizing, hinz2011seeding, zubcsek2011advertising, libai2013decomposing,lanz2019climb}.\footnote{We generally describe the goal as maximizing total adoption. Naturally, one might intervene to reduce an outcome, such as with vaccinations \citep{cohen2003efficient} or other preventative measures, as in the second empirical application below \citep{paluck2016changing}. Similar ideas also apply to allocating a budget for measurement in a network, rather than intervention, as with selecting nodes to serve as sensors for outbreak detection \citep{leskovec2007cost,christakis2010social}.} 

In this paper we show how to efficiently evaluate such targeting strategies from field experiments. Crucial to our approach is the insight that many strategies of interest are \emph{stochastic seeding strategies}, meaning that the targeted individuals are not selected deterministically but rather selected according to a probability distribution over eligible seed sets. This insight allows us to construct reweighting estimators using ideas largely adapted from the counterfactual policy evaluation and importance sampling literatures.  In our analysis these estimators sometimes provide more than a four-fold boost in effective sample size over simple difference-in-means estimators.  Given these estimators, we also show how to gain further precision via optimized experimental designs, and how to conduct off-policy evaluation of field experiments that were not explicitly designed to evaluate targeting strategies.

\begin{figure}[t]
\begin{minipage}{6in}
  \centering
  \raisebox{-0.5\height}{\includegraphics[height=9cm,angle=90]{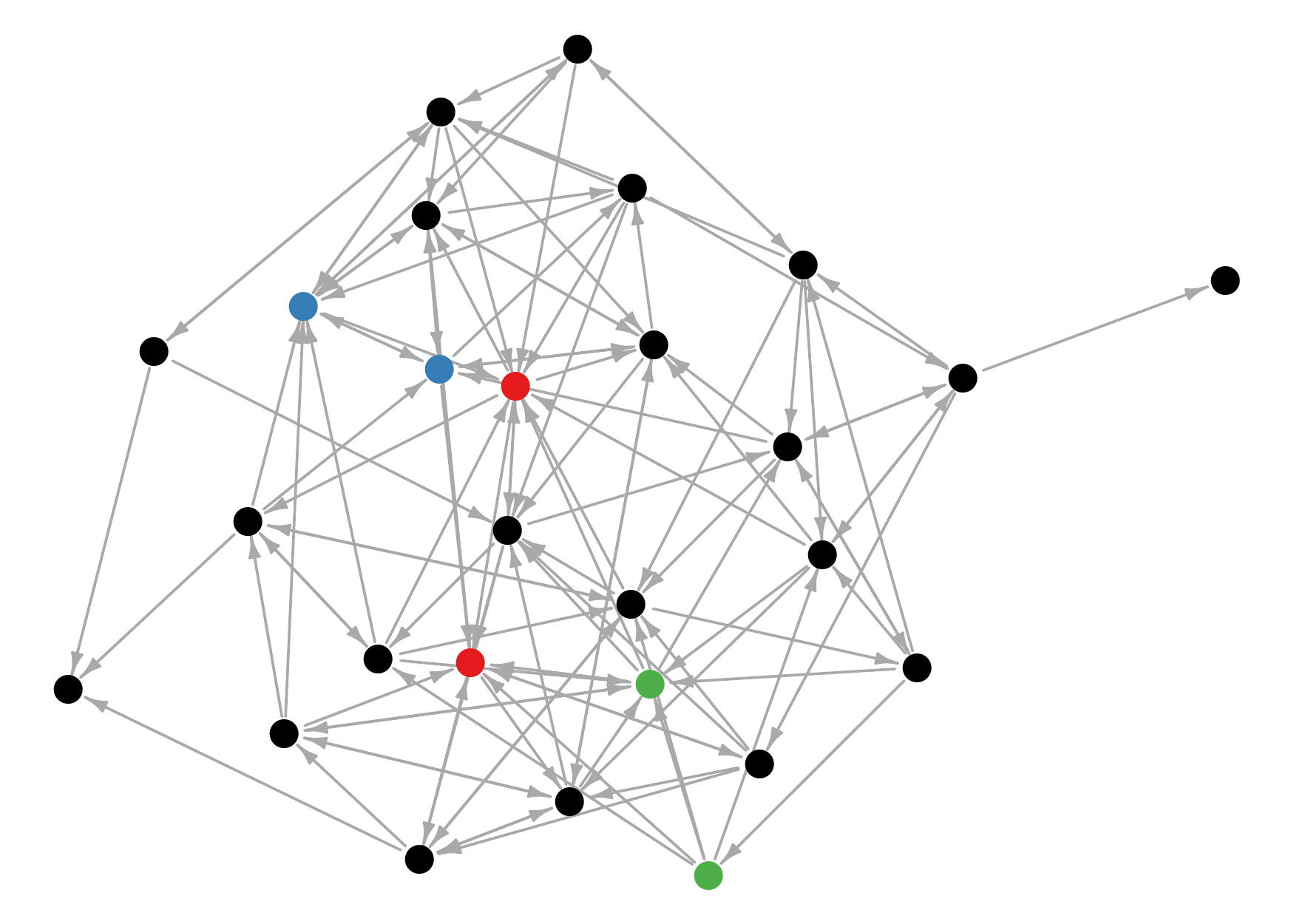}}
  \raisebox{-0.5\height}{\includegraphics[height=8cm]{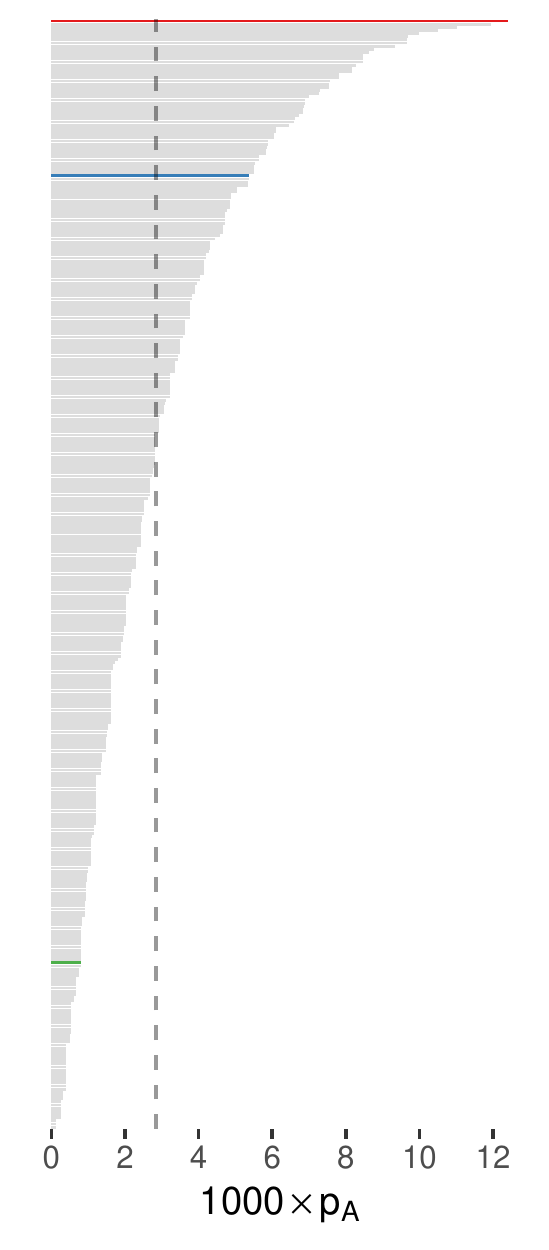}}
\end{minipage}
\caption{
Illustration of the probability of a seed set being selected under the one-hop seeding strategy with seed sets of size $k = 2$. (left) Network for one village in \citet{cai2015social}, with three possible seed sets highlighted: the seed sets with maximum probability under one-hop targeting (red) and two other example seed sets (blue, green).
(right) Probability under the one-hop strategy ($p_A$) of all seed sets, with the highlighted seed sets (red, blue, and green) corresponding to those in the network on the left. The dashed line represents the uniform probability of each seed set under random seeding. \label{fig:seeding}}
\end{figure}

Our analysis can be used to compare arbitrary stochastic seeding strategies considered over the same eligible seed sets, but for concreteness much of our discussion and analysis focuses on a particular strategy that we call \emph{one-hop targeting}.\footnote{In some literatures this idea is called ``acquaintance sampling''~\citep{cohen2003efficient,maiya2011benefits}. \citet{kim2015social} call this idea ``nomination'' targeting, though it typically does not involve individuals nominating a particular peer to be selected; rather they specify a set of people as friends, kin, etc., and then one such peer is selected at random.}  In one-hop targeting a seed set of $k$ nodes is assembled by first randomly selecting $k$ nodes and then randomly selecting one of their network neighbors as a seed. This strategy seeks to take advantage of the \emph{friendship paradox}, Feld's (\citeyear{feld1991your}) observation that the average node has fewer connections than the average neighbor.\footnote{We note that one-hop targeting is different from edge sampling, the original mechanism identified by \citet{feld1991your} as driving the friendship paradox. See \cite{lattanzi2015power} and \citet{kumar2018network} for further discussion of these differences.}  Because friends of randomly-selected individuals are likely to have more connections than the randomly-selected individuals themselves, this one-hop strategy ostensibly increases the chances of selecting ``influential'' seeds, at least as measured by their number of connections. See Figure~\ref{fig:seeding} for an illustration of the probabilities of different seed sets on a small network. This strategy, which is \emph{local} in the sense that it does not require observation of the entire network, is further motivated by the fact that observing or measuring the entire social network for a population can be impractical or prohibitively expensive [such that it might be better to spend that budget simply treating a few more nodes \citep[cf.][]{akbarpour2017diffusion}].

While one-hop targeting and close variants have received substantial attention and advocacy~\citep{cohen2003efficient,gallos2007improving,christakis2010social,flodgren2011local,maiya2011benefits,kim2015social,chami2017social,banerjee2017using,ben2019extended}, empirical evidence that it leads to greater adoption rates remains limited.  In a field experiment on a study population of 5,773 individuals in  32 villages in rural Honduras, \citet{kim2015social} compare one-hop targeting to \emph{random targeting}, a baseline strategy where seeds are selected uniformly at random without using network information.\footnote{The authors use a third strategy where they target nodes with maximum in-degree, which we do not address here.} The goal was to measure the effectiveness of these targeting strategies for two public health marketing interventions aimed at encouraging adoption of chlorine for water purification and multivitamins for micronutrient deficiencies.  Selected seed individuals were provided information and coupons for these interventions, and the overall coupon redemption rate was used as the primary outcome.  The authors reported no statistically significant results for the redemption of chlorine coupons but a 95\% confidence interval of a 6.9\% to 17.9\% increase in the redemption of multivitamin coupons, though these inferences rely on strong parametric and within-village independence assumptions that would seem to be violated by the influence processes posited to cause the observed differences.
Reflecting optimism about one-hop targeting but also the need for further empirical research, that team is conducting a larger, follow-up field experiment \citep{shakya2017exploiting}.

The \citet{kim2015social} study demonstrates that even for quite large field experiments aimed at studying one-hop targeting, statistical power and nonparametric inference remain a challenge.  Though the study contained 5,773 individuals, the nature of the intervention and village-level outcome means that the true effective sample size may be as small as 32 villages.
Since field experiments remain expensive and imprecise, better methods are needed.  

To see how our methods for evaluating stochastic strategies apply to the \citet{kim2015social} study, consider a seed set selected by a uniformly at random strategy. If the observed seed set has positive probability under one-hop targeting (which it will, assuming all nodes in the seed set have positive in-degree) then it also provides information about the outcome under the one-hop targeting distribution, regardless of the fact that it was selected under the uniformly at random strategy.  
In fact, because the strategies are stochastic, a seed set selected uniformly at random will sometimes even have higher in-degrees than a set selected by one-hop targeting. This reversal plays out in the \citet{kim2015social} study: of the eight villages that were assigned to one-hop and random targeting for the two different products, in three villages (37.5\%) the random seed set had a higher mean in-degree than the one-hop seed set (see Appendix~\ref{app:kim}). Using these villages in a simple difference-in-means comparison of the sort used in the analysis by~\citet{kim2015social} can then in fact lead to imprecise estimators. In this work we show how we can instead use potentially \emph{all} villages to learn about \emph{both} targeting strategies. As a methodological contribution with no clear precedent in the policy evaluation literature, we resolve how to conduct valid nonparametric asymptotic inference when comparing stochastic strategies, as well as Fisherian randomization inference for small samples.

Our methods are not just restricted to the analysis of experiments that were designed specifically for the purpose of comparing seeding strategies, such as the \citet{kim2015social} study.
In fact, our estimators can be applied to \emph{any} experiment that randomizes node treatment in a collection of observed or partially observed networks.  This observation opens the possibility of analyzing a much larger range of studies, as a number of recent experiments in networks study interventions that use random targeting, designed without any stated goal of evaluating the efficacy of other strategies. In this paper we analyze two such studies that used random targeting, bringing new empirical results into the conversation about one-hop targeting. In Section~\ref{sec:empirical} we reanalyze data from the farmer's insurance marketing experiment in China conducted by~\citet{cai2015social} and the peer conflict study in U.S.~schools conducted by~\citet{paluck2016changing}, both of which used random seeding.  

As a final point, for potential future studies with an explicit goal of evaluating the efficacy of targeting strategies, we also present ways to optimize the design of such experiments to improve inferential power. The optimization, evaluated in Section~\ref{sec:design_and_ess} with a detailed discussion in Appendix~\ref{app:opt-design}, is based on the idea that some seed sets will be much more informative for comparing strategies than others, and such seed sets can be identified before having run the experiment. For example, for any two stochastic strategies, a seed set that is equally probable under the two strategies does not shed light on the difference between the two strategies. 

As stated before, the framework we provide does not just apply to one-hop and random targeting but rather can be applied to evaluate the difference between any two stochastic seeding strategies with the same support.
For example, in two field experiments involving marketing a lottery and vaccines in 734 Indian villages, %213 + 521 + 
\citet{banerjee2017using} study a close variant of the one-hop strategy where they select seeds in villages randomly from the people named by a small random sample of households. Like \citet{kim2015social}, their analysis does not make use of the known distribution of seeds selected by this strategy, and the resulting inferences suffer from lack of statistical power and precision (e.g., all key contrasts with random seeding have $p$-values $> 0.02$ and confidence intervals ranging over orders of magnitude).
There are many natural stochastic strategies based on both local or non-local information.  One-hop and random targeting both assume no more than local knowledge of the network; other local-information stochastic strategies include multi-hop targeting, targeting random neighbors of a set with particular demographic traits, a mixed strategy that targets either a random node or a one-hop neighbor, or perhaps targeting both a random set {\it and} one of their random friends. The last example would be expected to behave favorably in domains governed by complex contagion \citep{centola2007complex}.
Recently, other stochastic seeding strategies that offer some theoretical guarantees have been proposed \citep{wilder2018maximizing,eckles2019seeding}; these have not yet been studied in field experiments.
In settings where the full network is available, a wide range of stochastic strategies admit themselves: one could target nodes randomly proportional to degree (which is roughly achieved by one-hop targeting), randomly proportional to some more sophisticated centrality measure, or using some randomized algorithm to, e.g., greedily construct independent sets \citep{luby1986simple}.
Relatively little empirical attention has been devoted to some of the strategies mentioned here, but we hope that our proposed methods help to stimulate further research in this area and lower the barrier to entry for running more powerful field experiments.

\subsection{Related work}

The present work is related to both prior work on targeting in social networks and more general work in statistical methodology. We briefly review key related work on the network mechanisms of peer effects, field experiments attempting to understand the role of network information in targeting, the literature focused on one-hop targeting, and the statistical literature on importance sampling and counterfactual policy evaluation in causal inference.

The literature on peer effects or ``social contagion'' is large, including adoption and use of products with direct~\citep[e.g.,][]{tucker2008identifying,ugander2012structural,eckles2016estimating} and indirect~\citep[e.g.,][]{nair2004empirical} network effects, and even those with plausibly absent network effects \citep[e.g.,][]{aral2009distinguishing}.
Social network information may be used for targeting in at least two ways. First, individuals may be indirectly affected by interventions received by others. Second, due to (dis)assortativity, social networks encode otherwise latent information about individuals \citep[e.g.,][]{mcpherson2001birds,hill2006network,currarini2010identifying,altenburger2018monophily}, so characteristics of network neighbors can be used for targeting even if there are no expectations of peer effects. 
Furthermore, there can be substantial heterogeneity in peer effects \citep{bakshy2012social,aral2012identifying,aral2014tie} and, in aggregate, in combination with network structure, large differences in the total influence of an individual's adoption \citep{bakshy2011everyone,iyengar2011opinion,yoganarasimhan2012impact,galeotti2017targeting}. 

There is a substantial literature on optimal seeding and approximations thereof given a known network and a model of how individuals are affected by the intervention and others' adoptions. In this literature on ``influence maximization'', computer scientists have developed algorithms for the problem of finding a set of $k$ seeds so as to maximize expected adoption under various models \citep{domingos2001mining,richardson2002mining}. This influence maximization problem is NP-hard \citep{kempe2003maximizing}, so much of that work \citep[e.g.,][]{chen2009efficient} is concerned with efficient algorithms for tractable approximations under various models of social influence.
It has been observed that the costs associated with causing an individual to adopt can vary \citep[cf.][]{bakshy2011everyone} and influence and susceptibility may be correlated in the network \citep{aral2013engineering,aral2018social}; thus, the most influential individuals may be the most difficult to induce to adopt. 
As a result, credibly evaluating the efficacy of seeding strategies net of all these possible complexities is very difficult without large-scale field experiments. Seeding strategies can be quite sensitive to deviations from the simple theoretical models sometimes used. 

Researchers in many disciplines have conducted empirical studies of seeding strategies in attempts to evaluate the role of network information \citep[e.g.,][]{sikkema2000outcomes,campbell2008informal,hinz2011seeding,kim2015social,chami2017diffusion,beaman2018diffusion,beaman2015can,banerjee2017using}, and we expand here on a few representative examples. 
\citet{banerjee2013diffusion} studied the diffusion of a microfinance product in villages in India, seeding deterministically at village ``leaders'', and found that villages where the leaders had higher centrality in the village social network saw higher adoption. 
\citet{chami2017diffusion} studied the adoption of de-worming drugs in villages in Uganda, seeding the adoption with two seeds per existing volunteer health workers in each village. They observed that villages where the seeds had higher clustering saw more adoption.
These studies are ambitious attempts to interrogate the value of network information in seeding, but the lack of randomization in the seeding strategies---either by using stochastic strategies or by evaluating multiple (possibly deterministic) strategies in a randomized trial at the village  level---impede causal inference related to seeding strategies.
There is a growing set \citep{kim2015social, banerjee2017using, beaman2018diffusion, beaman2015can} of randomized trials in which multiple villages, firms, or schools are assigned to seeding strategies. \citet{beaman2015can} studied the diffusion of a new agricultural technology in 200 villages in Malawi, with four treatment arms considering different deterministic seeding strategies. While the overall design (randomly blending four strategies) was stochastic, the deterministic design of the four constituent strategies means that at most four seed sets had positive probability of being selected in any given village. As such, this design cannot be used to evaluate the efficacy of other strategies with broader support in the space of seed sets (such as random targeting or one-hop targeting). More formally, the design does not satisfy our Assumption~\ref{assumption:positivity} (positivity) discussed in Section~\ref{sec:estimators} as enabling the analysis in our work.

Prior research on one-hop targeting has examined how it shifts the distribution of degree and centrality measures of seeds \citep[e.g.,][]{lattanzi2015power,kumar2018network}, compared with random targeting. Some work has analyzed resulting outcomes under assumed parametric models of contagion \citep{kumar2018can, chami2017social}. Ideas based on the friendship paradox (which motivates one-hop targeting) have also been applied beyond seeding to the related problems of outbreak detection~\citep{christakis2010social} and immunization \citep{cohen2003efficient,borgs2010distribute,gallos2007improving,chami2017social}.
The work on one-hop targeting is very focused on the promise posed by a seeding strategy that might harnesses network information to increase adoption without requiring full knowledge of the network (as is required by, e.g., the influence maximization perspective). Our work is also interested in this promise, and while our evaluation framework is more general than just studying the difference between one-hop targeting and random targeting, a key goal of our work is to contribute a more efficient causal inference framework for evaluating one-hop targeting.

Our proposed estimators are adaptations of estimators familiar from the literatures on importance sampling, counterfactual policy evaluation in bandit and reinforcement learning \citep[e.g.,][]{dudik_doubly_2014, li_unbiased_2011, precup2000eligibility, li_minimax_2014, swaminathan_counterfactual_2015, swaminathan2017off}, treatment rules \citep[e.g.,][]{hirano2009asymptotics, manski2004statistical, manski_identification_2007}, and dynamic treatment regimes \citep[e.g.,][]{robins1986new, murphy_marginal_2001, murphy2003optimal, hernan2006comparison}, including recent applications in marketing~\citep{hitsch2018heterogeneous,simester2020efficiently}. \citet{dudik_doubly_2014} and \citet{athey_efficient_2017} include reviews that span these multiple literatures.
Much of this related methodological work focuses on evaluating and learning deterministic policies, though some work considers extensions where stochastic policies are addressed. For example, \citet{murphy2003optimal} consider policies that simply assign treatment with some non-degenerate probability; \citet{munoz2012population} consider shifts to the distribution of a continuous or many-valued treatment, such as the number of hours spent in vigorous exercise or time from onset of symptoms to treatment. Stochastic policies are sometimes considered not because they are of interest \emph{per se}, but because they solve positivity problems that may exist for evaluating deterministic interventions \citep{munoz2012population,kennedy2018nonparametric}. In other cases, these policies may be of interest because policy-makers have limited control over treatment; this latter case is more similar to the present setting.

\subsection{Paper organization}

The remainder of the paper is structured as follows. In the next section, we begin with a detailed presentation of the one-hop seeding strategy. We then introduce the problem of studying stochastic seeding in more detail, including defining estimands that compare expected outcomes under different seeding strategies. In Section~\ref{sec:estimators}, we present estimators for those estimands, which are adaptations of estimators familiar from importance sampling, counterfactual policy evaluation, and dynamic treatment regime literatures. We also derive the variance of these estimators, which lets us consider how effective sample size depends on both the strategies being compared and the experimental design. Section~\ref{sec:simulations} presents a number of simulations, including of a simple experiment using the design from \cite{kim2015social}, but also of the benefits of optimal design. We then consider applications to data from field experiments in Section~\ref{sec:empirical}, evaluating stochastic seeding strategies in both the experiments of \citet{cai2015social} and \citet{paluck2016changing}. Section~\ref{sec:discussion} concludes with a discussion of how the framework in this paper can be used to further the study of seeding strategies. 

\section{Problem Formulation}
\label{sec:problem}

Suppose we are interested in estimating the difference in adoption between two strategies (i.e., policies): one-hop and random targeting. These two seeding strategies will be used throughout this work as the main examples of our analysis framework, though all components of the framework apply generically to two stochastic seeding strategies $A$ and $B$ (with the exception of localized discussions that are clearly described as being specific to random and one-hop targeting). We have data from an experiment in which each of $N$ villages (or, e.g., schools, firms), labeled $i = 1, \dots, N$, was randomly assigned to either random targeting or one-hop targeting.  For each village we also have access to a graph $G_i = (V_i, E_i)$, where $V_i$ is the node set and $E_i$ is the edge set, which records the social connections among the $n_i = |V_i|$ residents of village $i$.

For simplicity, we focus on cases where, for a given village $i$, the strategies in the experiment targeted seed sets of a fixed size $k_i$, though extension to random $k_i$ is straightforward. Let $\cS_i = \{s \subseteq V_i : |s| = k_i\}$ denote the collection of all such seed sets, which comprise the \emph{eligible seed sets} in village $i$. Let $S_i$ be a random variable that represents the seed set selected in village $i$, sampled from $\cS_i$.  Formally, each stochastic seeding strategy for village $i$ imposes a non-degenerate probability distribution for the random variable $S_i$.  Let $p_i^A$ denote the seed set probability distribution for one-hop targeting in village $i$, and let $p_i^B$ denote the seed set probability for random targeting in village $i$.  Throughout will use $A$ to represent one-hop targeting and $B$ to represent random targeting but again emphasize that much of our approach can be used to compare arbitrary stochastic strategies $A$ and $B$.

For every village the experiment produces a village-level response $Y_i$.  We follow the potential outcomes framework~\citep{neyman1923application,rubin1974estimating} and assume that the potential outcomes for each village are fixed at the seed set level, meaning that $Y_i = y_i(S_i)$, where $y_i: \cS_i \to \cY$ is a function mapping from the space of seed sets to the outcome space $\cY$.\footnote{Writing the potential outcomes for village $i$ as a function of only the seed set for village $i$ implies (a) the absence of interference between villages and (b) that the seeding strategy used only affects outcomes via which seed sets are selected (e.g., the strategy is not announced).
}
In most of the cases that we study, $Y_i$ is a count or fraction of adopters of the village and so $\cY = \mathbb{Z}$ or $\cY = [0, 1]$, but this fact is not important for any of our results.

Given the above notation, we can now state the estimand for the experiment in both a finite population and superpopulation framework.  The goal of the experiment is to estimate the difference in expected outcomes for each of the two targeting strategies:
\begin{equation}
\taufp = \avgin \left[\E_i^{A}[y_i(S_i)] - \E_i^{B}[y_i(S_i)]\right],
\label{eqn:taufp}
\end{equation}
where $\E_i^A$ and $\E_i^B$ denote expectation over $S_i \sim p_i^A$ and $S_i \sim p_i^B$, respectively.
Here $\taufp$ is a finite population estimand, which considers the villages to be fixed.

We also consider estimation of the superpopulation estimand, in which the villages (and the corresponding networks $G_i$) are viewed as an i.i.d.~sample from an infinite superpopulation.  In this case, we may consider a single \emph{design distribution} $p_\Delta$, from which seed sets are sampled i.i.d.  The goal is to study the difference between the superpopulation one-hop targeting distribution (denoted $p_A$) and the superpopulation random targeting distribution (denoted $p_B$).  The finite population distributions discussed above, $p_i^A$ and $p_i^B$, result from conditioning $p_A$ and $p_B$ on the observed network $G_i$ for each sampled village $i$.

In the superpopulation framework seed sets are i.i.d.~draws of a random variable $S$. We observe i.i.d.~realizations of a response variable $Y = y(S)$, and the superpopulation estimand is then
\begin{equation}
\tausp = \E_A[y(S)] - \E_B[y(S)],
\label{eqn:tausp}
\end{equation}
where $\E_A$ and $\E_B$ denote expectation with respect to the superpopulation targeting distributions.
Notice that $\tausp$ can be also be written as
\begin{align}
\label{eqn:tausp2}
\tausp = \E_\Delta\left[\frac{p_A(S) - p_B(S)}{p_\Delta(S)}y(S)\right].
\end{align}

\section{Estimators}
\label{sec:estimators}
We now derive estimators for the above estimands. Let $Z_i$ be the village-level treatment indicator, where $Z_i = 1$ if village $i$ is assigned to one-hop targeting and $Z_i = 0$ if village $i$ is assigned to random targeting.  The resulting data consist of the realized treatment assignments $z_1, \dots, z_N$, seed sets $s_i$, and outcomes $y_1, \dots, y_N$.  We use lowercase letters to indicate that these are observed values.

The simplest estimator is the difference-in-means estimator, which is simply the difference in sample means for villages assigned to each seeding strategy:
\[\DM = \frac{1}{\sumin z_i} \sumin z_iy_i - \frac{1}{\sumin(1 - z_i)}\sumin(1 - z_i)y_i.\]
The difference-in-means estimator is unbiased for both $\taufp$ and $\tausp$, but we can increase precision~\citep{sarndal1976uniformly} by noting that each observation is potentially informative about multiple targeting strategies: if an observed seed set has positive probability under a particular seeding strategy, then it provides information about that strategy.

We assume that the experiment selects $Z_i \sim \on{Bernoulli}(\rho)$, where $\rho \in (0, 1)$ is the treatment assignment probability.  Each seed set $S_i$ is then sampled from the mixture distribution
\[S_i \sim \rho p_i^A + (1 - \rho)p_i^B.\]
We refer to this distribution as the {\it design distribution} for village $i$, denoted by $p_i^\Delta$, with probabilities given by
\begin{equation}
\P_i^\Delta(S_i = s) = \rho \P_i^A(S_i = s) + (1 - \rho)\P_i^B(S_i = s).
\label{eqn:design}
\end{equation}

Since $p_i^A$, $p_i^B$, and $p_i^\Delta$ are all completely known, we in particular know the exact probabilities corresponding to the observed seed sets.  Let
\begin{align*}
a_i &= \P_i^A(S_i = s_i), \\
b_i &= \P_i^B(S_i = s_i), \\
d_i &= \P_i^\Delta(S_i = s_i),
\end{align*}
denote the corresponding observed probabilities. 
We can then compute reweighting estimators by defining the weights
\begin{equation}
\label{eqn:weights}
w_i^A = a_i / d_i, \qquad w_i^B = b_i / d_i.
\end{equation}
and then taking
\begin{equation}
\hat \tau = \avgin (w_i^A - w_i^B) y_i.
\label{eqn:ht}
\end{equation}
Comparing $\hat \tau$ to $\tausp$ in equation~\eqref{eqn:tausp2}, we can see that $\E_{\Delta} [ \hat \tau ] = \tausp$; a full proof is given as part of the proof of Proposition~\ref{prop:ht} in Appendix~\ref{app:proofs}. This estimator is an analog of the standard Horvitz--Thompson estimator~\citep{horvitz1952generalization} used in survey sampling for estimating population quantities.  In that setting, if $a_i$ and $d_i$ are the sample and population populations for stratum $i$, then $N^{-1} \sumin a_i / d_i$ is an unbiased estimate of the population mean. 
Alternatively, we may define the normalized weights
\[\tilde w_i^A = \frac{w_i^A}{N^{-1}\sum_j w_j^A}, \qquad \tilde w_i^B = \frac{w_i^B}{N^{-1}\sum_j w_j^B},\]
and obtain an analog of the H\'ajek estimator
\begin{equation}
\label{eqn:hajek}
\tilde \tau = \avgin (\tilde w_i^A - \tilde w_i^B) y_i.
\end{equation}

 It is important to note that while we borrow the idea of inverse probability weighting here, equations~\eqref{eqn:ht} and \eqref{eqn:hajek} are both differences of two correlated terms and so theoretical results about these estimators do \emph{not} follow from standard results in survey sampling. Indeed, the correlation between the two terms in (e.g.) equation~\eqref{eqn:ht} is crucial, as it captures our principal idea that a single village carries information about both strategies.  This is discussed further in Section~\ref{sec:inference} on asymptotic inference.

The standard versions of these estimators are also familiar in the importance sampling literature, where the Horvitz--Thompson estimator is known as an \emph{unnormalized importance sampling estimator} and the H\'ajek estimator is known as an \emph{self-normalized importance sampling estimator} (see, for example, \citet{mcbook}).  In that context, data is provided by means of an \emph{importance distribution} but the desired quantities are population moments of a different \emph{reference distribution}.  For our application, the design distribution, $p_i^\Delta$, serves as the importance distribution and the two targeting distributions, $p_i^A$ and $p_i^B$, are the reference distributions.

In the importance sampling literature there are competing arguments for whether to use the unnormalized estimator $\hat \tau$ or the normalized estimator $\tilde \tau$, and the optimal choice will depend on the particular application.  See, for example, the discussion in~\citet[Ch. 9]{mcbook}.  The Horvitz--Thompson estimator is unbiased, as the unnormalized weights $w_i^A$ and $w_i^B$ have mean one.  However, if they have excessive variance then the resulting Horvitz--Thompson estimator will be quite unstable.  The  H\'ajek estimator remedies this problem by forcing the mean of the weights $\tilde w_i^{A}$ and $\tilde w_i^{B}$ to be exactly equal to one.  The H\'ajek estimator is biased, but this bias is negligible in large sample sizes.  In our case, since the seed set probabilities are usually extremely small, we generally expect the self-normalized H\'ajek estimator to be more precise.\footnote{
Note that the H\'ajek estimator is not a regularized estimator that shrinks towards zero; rather the bias arises from the presence of a random variable in the denominator, as with estimation of ratios.
}

\subsection{Positivity}

We require that the design satisfies a contrastive notion of positivity, restricted to seed sets where the probabilities differ, in order for the observed data to provide the necessary information about the two targeting strategies.
\begin{assumption}[Positivity]
\label{assumption:positivity}
For every village $i$ and seed set $s_i \in \cS_i$, if $p_i^A(s_i) \neq p_i^B(s_i)$, then $p_i^\Delta(s_i)>0$.
\end{assumption}
\noindent
Positivity is automatically satisfied for the mixture distribution considered in equation~\eqref{eqn:design}, but the researcher may also consider other designs such as the optimized designs discussed in Appendix~\ref{app:opt-design}.  Assumption~\ref{assumption:positivity} is needed for off-policy evaluation as well. For example, suppose a seeding experiment on a collection of networks was designed for another purpose and used completely random assignment to treatment in all networks.  In this case the design distribution is simply the random targeting distribution, $p_\Delta = p_B$, rather than the mixture distribution given by equation~\eqref{eqn:design}, and Assumption~\ref{assumption:positivity} reduces to requiring $p_i^B(s_i) > 0$ whenever $p_i^A(s_i) > 0$, which is the much more standard positivity assumption used in general importance sampling and causal inference \citep{imbens2015causal} settings. Then $w_i^B = b_i / b_i = 1$, so the random targeting mean is estimated using the standard sample mean of the observations, and the off-policy estimate for policy $A$ is obtained by using the weights $w_i^A = a_i / b_i$.

We make two remarks on implications of our positivity assumption in more complex designs.  

\begin{remark}
In some common cases, Assumption \ref{assumption:positivity} may not be strictly satisfied for the stochastic targeting strategies we have considered so far. For example, an experiment may have used a design that blocked (i.e., pre-stratified) on observables (e.g.,~household income), such that is was impossible for all $k$ seeds to be, e.g., households in the highest income category. One can then consider variations on the stochastic seeding strategies that condition on, e.g., the relevant balance of observables among seeds or between seeds and non-seeds.
\end{remark}

\begin{remark}
Experimental designs may often be mixtures of stochastic and deterministic strategies (e.g., \citet{kim2015social} use a mixture of random seeding, one-hop seeding, and selecting the maximum in-degree nodes). The unconditional design distribution $p_i^\Delta$ may still satisfy Assumption \ref{assumption:positivity} even if not all, or even none, of the component distributions do so individually.
\end{remark}

\subsection{Computing seed probabilities}
\label{sec:computingprobs}

Using a Horvitz--Thompson or H\'ajek estimator to compare stochastic strategies requires being able to efficiently compute or estimate the seed set probabilities for the strategies being compared. 
For basic random targeting, these probabilities are straightforward: all eligible seed sets are equally likely. When  policy $B$ is random targeting then $p_i^B$ is characterized by the uniform probabilities
\begin{equation}
\label{eqn:rand-prob}
\P_i^{B}(S_i = s) = \binom{n_i}{k}^{-1}, \qquad \text{for all }s \in \cS_i, \text{ } i=1,\ldots,n,
\end{equation}
where $\P_i^B$ denotes probability with respect to $p_i^{B}$. Notice that this probability is independent of the network structure of village $i$, depending only on the number of nodes $n_i$.

For more complicated targeting strategies, computing the seed set probability can be challenging. We first examine the case of one-hop targeting, where we describe an exact dynamic program for computing the seed set probabilities, explained in more detail in Appendix~\ref{app:probs}. This dynamic programming solution is somewhat general; it can be adapted to any network-aware stochastic policy that decomposes into node-level seed selection probabilities that are themselves tractable. We then also consider a general Monte Carlo estimation procedure for the probabilities under more complex strategies with dependence between nodes, expanded on in Appendix~\ref{app:probs}, with a potentially useful variance bound.

The one-hop targeting probabilities for village $i$ depend on the structure of the graph $G_i$. 
There are several minor variations on one-hop targeting, depending on how one chooses to operationalize the seeding instructions. Are ``nominator'' individuals selected with replacement? Are seeds selected with replacement? In this work, we assume nominators are selected with replacement---not doing so requires considering the sequential nature of how the social network slightly changes from nomination to nomination---but seeds are selected without replacement.
By selecting nominators with replacement, draws of seed individuals become independent. Selecting seeds without replacement guarantees that there will be $k$ unique nodes in the eventual seed set.

To compute seed set probabilities under this one-hop seeding strategy we first consider the case in which seeds themselves are also drawn with replacement (meaning seed sets are in fact multisets, possibly containing multiple copies of a single seed). 
Let $\P_i^{A,\text{repl}}$ denote the probability with respect to $p_i^A$ with replacement, where strategy $A$ is this one-hop targeting policy. 
For $k = 1$, the probability of selecting an individual node $v$ is simply
\begin{equation}
\label{eqn:nom-prob1node}
\P_i^{A} \left (S_i = v \right )  =  \frac{1}{n_i} \sum_{u \in \mathcal N_{\text{in}} (v)} \frac{1}{d_u^{\text{out}}},
\end{equation}
where $\mathcal N_{\text{in}} (v)$ denotes the set of in-neighbors of $v$, and $d_u^{\text{out}}$ denotes the out-degree of node $u$. 
Then $\P_i^{A,\text{repl}}$ for unordered sets with $k>1$ nodes becomes:
\begin{align}
\nonumber
\P_i^{A,\text{repl}}(S_i = s) 
&= k! \prod_{v \in s} 
\P_i^{A} \left (S_i = v \right )  \\
&= k! \prod_{v \in s}
\frac{1}{n_i} \sum_{u \in \mathcal N_{\text{in}} (v)} \frac{1}{d_u^{\text{out}}}.
\label{eqn:nom-prob}
\end{align}
The $k!$ in the above expression comes from the fact that we seek the probabilities for unordered sets. For a given seed set $s$ this probability with replacement is straight-forward to compute.

With the above probabilities in hand, we can now translate each probability with replacement to one without replacement (where the seeds are unique). We can do this translation if we know the overall probability of selecting a set of size $k$ that is unique. In $G_i$ there are $\binom{n_i}{k}$ unique unordered sets of size $k$, and when this quantity is manageable we can simply compute the total probability 
\begin{equation}
\label{eqn:nom-sum}
\pi_i = \sum_{s \in \mathcal S_i : s \text{ unique}} \P_i^{A,\text{repl}}(S_i = s),
\end{equation}
which lets us use a simple normalization to compute the one-hop targeting probabilities without replacement:
\begin{equation}
\label{eqn:nom-prob2}
\P_i^A(S_i = s) = \frac{1}{\pi_i} \P_i^{A,\text{repl}}(S_i = s) .
\end{equation}

When the network size $n_i$ is modest and $k$ is very small (e.g., $k=2$ or $3$), we can simply enumerate the subsets of size $k$ exhaustively and compute this sum in equation~\eqref{eqn:nom-sum} directly. However, of the villages in the \cite{cai2015social} data that we analyze, one example village has 49 nodes and 13 seeds, meaning that there are $\binom{49}{13} \approx 262$ billion unique seed sets. In such settings (or in larger networks, such as those in the \cite{kim2015social} experiment), we require a more sophisticated procedure for computing $\pi_i$. In Appendix~\ref{app:probs} we show that for the one-hop strategy considered here, and indeed any stochastic policy that decomposes into independent node-level probabilities, the total probabilities $\pi_i$ can be computed efficiently via a dynamic program. We employ this dynamic program in our analyses in Section~\ref{sec:empirical}. 

In Appendix~\ref{app:probs} we also give a Monte Carlo estimation procedure for the total probability $\pi_i$ for generic strategies that can be much more complicated than one-hop seeding. This estimation procedure requires only that the set probabilities with replacement are computable, even if they don't decompose cleanly into node-level probabilities, and thus makes it possible to use Horvitz--Thompson and H\'ajek estimators to compare much more complicated strategies than just one-hop and random seeding. We also provide a simple upper bound on the variance of the Monte Carlo estimator in terms of the minimum and maximum probability of any given set being selected.

\subsection{Asymptotic inference}
\label{sec:inference}

In Section~\ref{sec:problem} we described how counterfactual evaluation of village-level outcomes is possible for non-deterministic targeting strategies.  The nature of our problem makes standard Neyman-style variance estimates~\citep[cf.][]{aronow2013class,imbens2015causal} for the finite population average  treatment effect $\taufp$ problematic, shifting our attention to the superpopulation average  treatment effect $\tausp$.  

The difficulties in the finite population setting stem from the fact that the observations are drawn from seed sets of arbitrarily different sample spaces $\cS_i$.  As a result, observations from village $i$ provide no information about village $j$, even if $i$ and $j$ were exposed to the same treatment strategy.  For example, consider the Horvitz--Thompson estimator $\hat \tau$ defined in equation~\eqref{eqn:ht}.  Since villages are independent, $\hat \tau$ has variance
\[\var(\hat \tau) = \frac{1}{N^2} \sumin \var[(w_i^A - w_i^B) y_i].\]
The village-level variances must be estimated separately, and because we observe only a single seed set in each village $i$, estimating any of the terms $\var[(w_i^A - w_i^B) y_i]$ is impossible.  Therefore, we focus on inference for the superpopulation average treatment effect $\tausp$, so that information can be combined across different villages.

In the standard importance sampling problem, the goal is to estimate a single population mean.  Our problem differs in that the importance sampled data are repurposed to estimate a difference of two population means, rather than a single population mean.  These estimates are correlated, so we cannot rely on standard importance sampling variance expressions and variance estimates from the literature.  In what follows we compute novel expressions for the variances and variance estimates for two generic strategies $A$ and $B$ (not just one-hop and random targeting). Proofs of these two propositions appear in Appendix~\ref{app:proofs}.

\begin{proposition}
Let $S \sim p_\Delta$ be a random seed set and let $P_A = p_A(S)$, $P_B = p_B(S)$, and $P_\Delta = p_\Delta(S)$ be random variables representing the probabilities corresponding to seed set $S$.  Let $Y = y(S)$, $W_A = P_A / P_\Delta$, and $W_B = P_B / P_\Delta$.  Then under Assumption~\ref{assumption:positivity} the Horvitz--Thompson estimator $\hat \tau$, defined in equation~\eqref{eqn:ht}, has expectation $\E[\hat \tau] = \tausp$ and $\var(\hat \tau) = V_{\hat \tau} / N$, where 
\begin{equation}
\label{eqn:ht-variance}
V_{\hat \tau} = \E \left[\frac{1}{P_\Delta^2}\left((P_A - P_B)Y - \tausp P_\Delta\right)^2\right] = \E[ ((W_A - W_B)Y - \tausp)^2].
\end{equation}
\label{prop:ht}
\end{proposition}
It is easy to construct an unbiased estimate of the variance expression in Proposition~\ref{prop:ht} by substituting in observed sample quantities, producing the variance estimator
\begin{equation}
\label{eqn:ht-var-est} a
\hat V_{\hat \tau} = \avgin \frac{1}{d_i^2}((a_i - b_i)y_i - \hat \tau d_i)^2 = \avgin ((w_i^A - w_i^B) y_i - \hat \tau)^2.
\end{equation}
Asymptotic normality of the Horvitz--Thompson estimator is immediate from Lindeberg's condition for a central limit theorem.

The H\'ajek estimator is not unbiased, but it is correct in large samples in the sense that it is consistent and asymptotically normal.
\begin{proposition}
\label{prop:hajek}
Let $S$, $P_A$, $P_B$, $P_\Delta$, and $Y$ be as in Proposition~\ref{prop:ht}.  Let $\mu_A = \E_A[Y]$ and $\mu_B = \E_B[Y]$.  Then under Assumption~\ref{assumption:positivity} the H\'ajek estimator $\tilde \tau$, defined in equation~\eqref{eqn:hajek}, satisfies $\tilde \tau \to \tausp$ as $N\to\infty$ and 
\[\sqrt{N}(\tilde \tau - \tausp) \wto \mathcal{N}(0, V_{\tilde \tau}),\]
where
\begin{equation}
\label{eqn:hajek-variance}
V_{\tilde \tau} = \E \left[\frac{1}{P_\Delta^2} \bigg(\mu_A P_A - \mu_B P_B - Y(P_A - P_B)\bigg)^2\right].
\end{equation}
\end{proposition}

We can estimate the H\'ajek variance with
\begin{equation}
\label{eqn:hajek-var-est}
\hat V_{\tilde \tau} = \avgin \frac{1}{d_i^2}(\tilde \mu_A a_i - \tilde \mu_B b_i - y_i(a_i - b_i))^2,
\end{equation}
where $\tilde \mu_A = N^{-1} \sumin \tilde w_i^A y_i$ and $\tilde \mu_B = N^{-1} \sumin \tilde w_i^B y_i$ are the H\'ajek plug-in estimates of the population means.

Lastly, when we consider the difference-in-means estimator $\hat \tau_{DM}$ we use the standard Neyman conservative variance estimator \[\hat V_{\hat \tau_{DM}} = \frac{S_1^2}{N_1} + \frac{S_0^1}{N_0},\]
where $S_1^2, S_0^2$ are the within-group sample variances and $N_1, N_0$ are the group sample sizes.

These variance estimators can be used to construct confidence intervals and conduct hypothesis tests using normal theory.

\subsection{Exact tests via Fisherian randomization inference}
\label{sec:fish}
The preceding variance estimators and their use for inference rely on asymptotic theory. Measures of effective sample size (see Appendix~\ref{app:ess}), or other diagnostics, may caution against relying exclusively on such approximations. We may instead wish to conduct exact finite-sample inference without relying on parametric assumptions. We thus briefly consider Fisherian randomization inference \citep{fisher_statistical_1925, fisher_design_1935} for $\tausp$ and $\taufp$. Typically, null hypotheses would posit zero average effects, whether in the finite population, $H_0: \taufp = 0$, or in a superpopulation, $H_0: \tausp = 0$. Using Fisherian randomization inference, we more readily test a sharp null hypothesis that outcomes are not affected by the seed set,
\begin{equation}
\label{eqn:sharp_null}
H_0^\text{sharp}: y_i(s) = y_i(s') \text{ for all } s, s' \in \cS_i, i \in \{1, ..., n\},
\end{equation}
or a similar hypothesis for a superpopulation. We can conduct an exact test of $H_0^\text{sharp}$ via Fisherian randomization inference by drawing counterfactual seed sets according to the design and computing a test statistic with these counterfactual seed sets and the observed outcomes, as the outcomes are unchanged under $H_0^\text{sharp}$.

While such a test is exact (i.e., results in no greater than the nominal Type I error rates under $H_0^\text{sharp}$) with any choice of test statistic, it is common to use a ``Studentized'' test statistic, which can yield tests that are also asymptotically valid under a non-sharp null $H_0$ for the superpopulation estimand \citep{chung2013exact} or finite population estimand \citep{wu2018randomization,cohen2020gaussian}; for the H\'ajek estimator, this Studentized test statistic is $\tilde{\tau} / \sqrt{\hat{V}_{\tilde \tau}}$. Thus, a test using such a test statistic is interpretable as informative about $\taufp$ and $\tausp$, though this would then again rely on, perhaps refined, asymptotic approximations.

\section{Simulations}
\label{sec:simulations}
In order to study our estimators in a setting where we can observe counterfactual outcomes, in this section we run simulations of behavior spreading on village networks according to a known model. We study the accuracy of the variance estimates and resulting coverage rates, the feasibility of off-policy evaluations, and a comparison of a commonly used experimental design for comparing seeding strategies vs.~an optimized design.
 
In order to accurately capture the network structure and heterogeneity exhibited among villages, we use the networks from~\citet{cai2015social}, the same networks we study in the empirical analysis of actual insurance decisions in Section~\ref{sec:empirical}. See the empirical analysis for a discussion of our pre-processing steps for that data, which are less relevant to the present simulations.  See also Table~\ref{table:cai-summary-stats} in Section~\ref{sec:empirical} for a list of summary statistics of that data.

\subsection{Performance in simple designs}
\label{sec:simulations-simple}
In this simulation, villages are assigned to one-hop or random targeting using Bernoulli$(0.5)$ coin flips. This is similar to the design in \citet{kim2015social}, but without blocking by village characteristics for simplicity. We fix the seed sets for all interventions to $k = 2$.  

To generate outcomes, we use a model with endogenous social interactions such that latent utilities are  linear-in-means.  Our model is a dynamic model similar to that used in~\cite{eckles2017design}, which can be regarded as a noisy myopic best response model in a semi-anonymous game with strategic complements. 
Let $S_{ij}$ be an indicator for whether individual $j$ in village $i$ is selected as a seed individual. Let $Y_{ij,t} \in \{0,1\}$ denote the adoption state of individual $j$ in village $i$ at time $t$. We set the initial set of adopters to be the seed sets, $Y_{ij,0} = \one\{S_{ij} = 1\}$.
  Then we define the $t$-th time step response using the probit model
\begin{align}
Y_{ij,t}^* &= \alpha + \beta Z_{ij,t} + \ep_{ij,t}, \label{eqn:sim-model}\\
Y_{ij,t} &= \max\{Y_{ij,t-1},\one(Y_{ij,t}^* > 0)\}.\nonumber
\end{align}
The intercept $\alpha$ captures a baseline threshold for adoption.
Let $G_{ijk}$ denote the $jk$-entry of the adjacency matrix for the network of village $i$. Let $d^{-}_{ij} = \sum_k G_{ijk}$ and $d^{+}_{ij}  = \sum_j G_{ijk}$ be the out- and in- degrees of individual $j$ in village $i$. 
Define $\tilde{G}_{ijk} = G_{ijk} / d^{+}_{ij}$ as that entry in the row-normalized adjacency matrix if $d^{+}_{ij} > 0$ and zero otherwise.
Then for time step $t$, we let 
\[Z_{ij,t} = \sum_k \tilde{G}_{ijk} Y_{ik,t-1} \]
be the mean of neighboring responses of the previous time step.  The parameter $\beta$ thus captures the endogenous social effect portion of the utility linear-in-means model. 

We use independent $\ep_{ij,t} \sim \mathcal{N}(0, 1)$ noise, which is homoscedastic across time and individuals.  The linear response $Y_{ij,t}^*$ is then dichotomized at zero.  We also require that $Y_{ij,t} = 1$ if $Y_{ij,t-1} = 1$, which enforces the constraint that adopters cannot revert to a state of non-adoption.  The village-level response $Y_i$ is the fraction of adopters after a maximum number of time steps $T$ have been completed, $Y_i = n_i^{-1} \sum_{j = 1}^{n_i} Y_{ij, T}$. In this model, all individuals eventually adopt, since there is an independent, positive probability of adoption in each time step. We set $T = 3$, noting that the average pairwise distance within most villages is less than $3$.

For parameter values, we vary $\alpha \in \{-3, -2, -1, 0\}$ and $\beta \in \{ 0, 2, 4, 6, 8, 10\}$.
For each of 5,000 Monte Carlo replicates, we consider the following simulation procedure.  
We conduct a simulated experiment by sampling $N \in \{50, 150\}$ villages without replacement\footnote{Simulations sampling villages with replacement, which more directly maps to consideration of a (trivial) superpopulation estimand, do not essentially modify any results.}
and assigning each village to either one-hop targeting or random targeting via a Bernoulli$(\rho)$ random variable (with $\rho = 1/2$) and compute difference-in-means, Horvitz--Thompson, and H\'ajek estimators as well as the corresponding variance estimates described in Section~\ref{sec:inference}.
The estimand is calculated by simulation by taking the difference in adoption rates when all villages are assigned to one-hop targeting, compared to random targeting.

In Appendix \ref{app:simulations} we present variations on these simulations, including some with another widely-used model of contagion (an independent cascade or SIR model). The results are qualitatively consistent with the results under the above model.

\subsubsection*{Results}

\begin{figure}[bt]
\centering
\includegraphics[width=\textwidth]{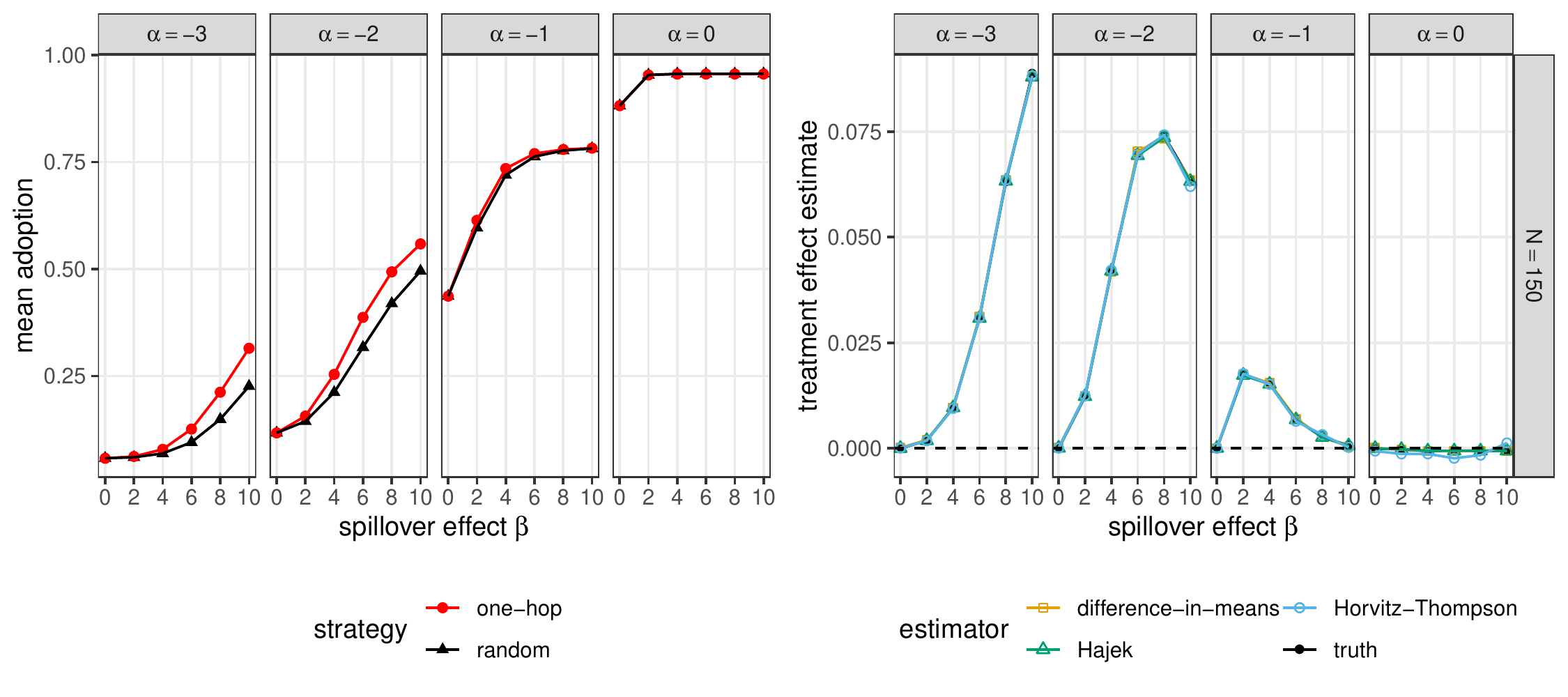}
\caption{(left) True mean adoption rates for random and one-hop targeting for the simulation setup described in Section~\ref{sec:simulations-simple}. (right) True treatment effect and estimated values from the difference-in-means, Horvitz--Thompson, and H\'ajek estimators.  Columns vary the intercept $\alpha$. The horizontal axis varies the spillover effect $\beta$.}
\label{fig:sim-true-tau}
\end{figure}

\begin{figure}[bt]
\centering
\includegraphics[width=\textwidth]{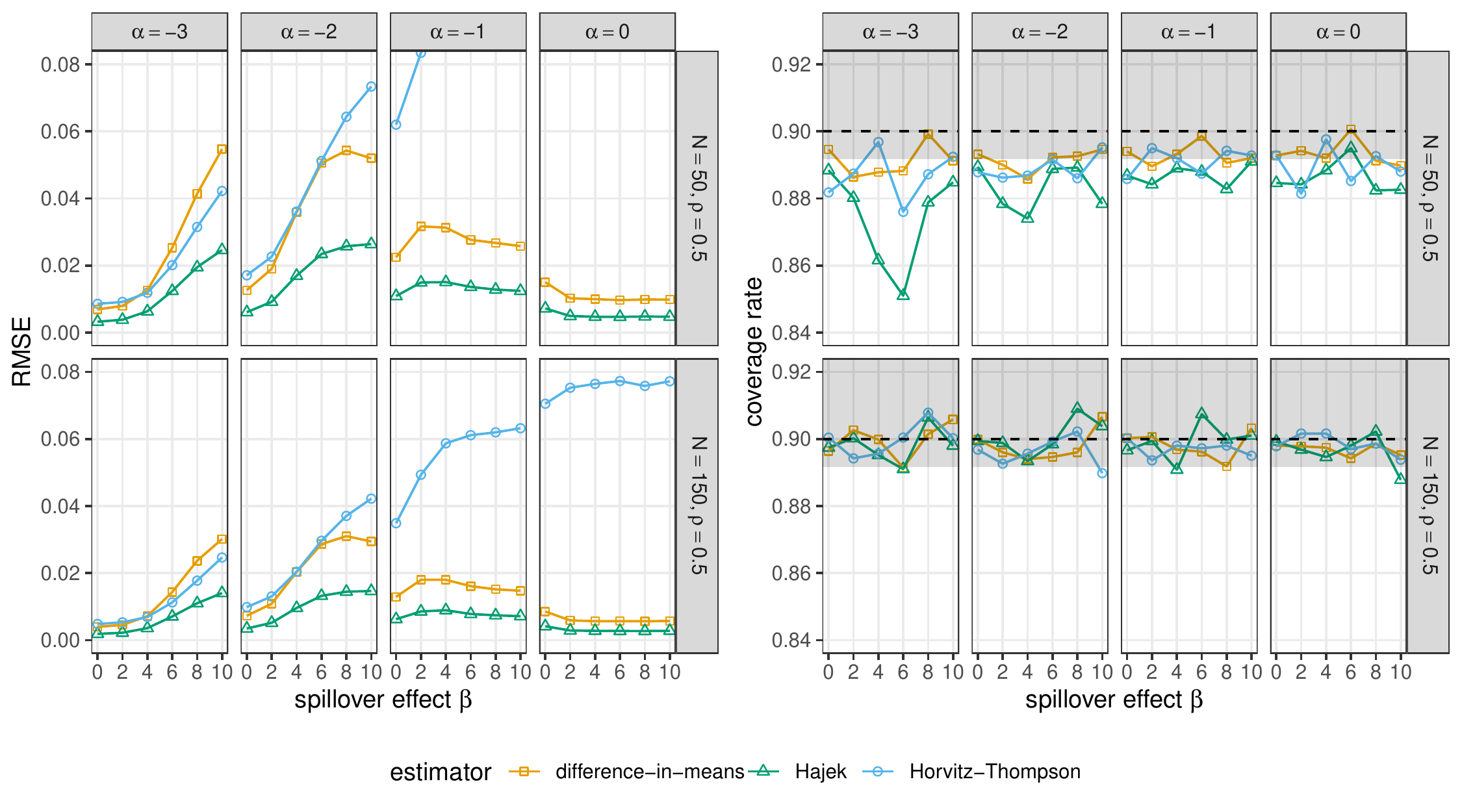}
\caption{(left) Root mean-squared-error of estimators in the simulation setup described in Section~\ref{sec:simulations-simple} (i.e., the Bernoulli design with $\rho = 1/2$ villages assigned to one-hop in expectation). Columns vary the intercept $\alpha$, while rows vary the number of villages sampled. The horizontal axis varies the spillover effect $\beta$.
Error increases with $\beta$ as it produces more within-village dependence. The Horvitz--Thompson estimator has very high variance in some cases and so is not visible in some panels.
(right) Coverage rates for a 90\% nominal confidence interval; shaded area is the 95\% acceptance region ($p > 0.05$) for coverage being at least the nominal rate. All estimators have approximately nominal coverage with $N=150$, while with $N=50$ inference using this asymptotic approximation is anti-conservative, notably for small sample sizes with a very rare outcome ($\alpha = -3, N = 50$).
}
\label{fig:sim-rmse-coverage}
\end{figure}

Figure~\ref{fig:sim-true-tau} (left) displays the true mean adoption rates under random and one-hop targeting, where we broadly see a higher adoption rate from one-hop targeting under this model, though
the parameter values used result in substantial variation in adoption rates and treatment effects.
Figure~\ref{fig:sim-true-tau} (right) displays the average estimates along with the true treatment effect.
As all are approximately unbiased, these estimators are evaluated first according to their total error (Figure~\ref{fig:sim-rmse-coverage}, left). As expected, the Horvitz--Thompson estimators suffer from imprecision. On the other hand, the H\'ajek estimator generally has substantially lower error than the difference-in-means.
We also evaluate the variance estimators via the coverage of the asymptotic confidence intervals (Figure~\ref{fig:sim-rmse-coverage}, right). The coverage is generally approximately at the nominal rate for $N = 150$, though is sometimes significantly lower for all estimators when $N = 50$.

We can ask how much the reduced error apparent in Figure~\ref{fig:sim-rmse-coverage} (left) translates into increased statistical power for the proposed estimators. Across all settings, the H\'ajek estimator has a true standard error that is 39\% to 55\% smaller; this corresponds to the gains from collecting data from 172\% to 398\% more villages.
Figure~\ref{fig:sim-power} in Appendix~\ref{app:simulations} presents evaluation of the power of the simulation experiments (the fraction of experiments in which the null was rejected). For the response model and parameters used in our simulations, the H\'ajek estimator generally has substantially more power than the difference-in-means estimator, while the Horvitz--Thompson estimator is underpowered due to excessive variance. 

\subsection{Performance in off-policy evaluation}
\label{sec:simulations-off-policy}
Part of the promise of our proposed approach is the ability to use data collected from one stochastic seeding strategy, e.g.~random seeding, to evaluate other stochastic seeding strategies. We now repeat the previous simulations, but with a different experimental design: all villages are assigned to random seeding. That is, rather than assigning each village to either one-hop targeting or random targeting via a Bernoulli$(\rho)$ random variable with $\rho = 1/2$, $\rho$ is zero. The difference-in-means estimator is not defined for this design.

\subsubsection*{Results}

As with the previous design, the H\'ajek estimator does not exhibit substantial bias, except in the case of a rare outcome with a small sample size (e.g., $\alpha = -3, N = 50$), where the effective sample size becomes very small.
We thus again evaluate these estimators according to their total error. 

\begin{figure}[bt]
\centering
\includegraphics[width=\textwidth]{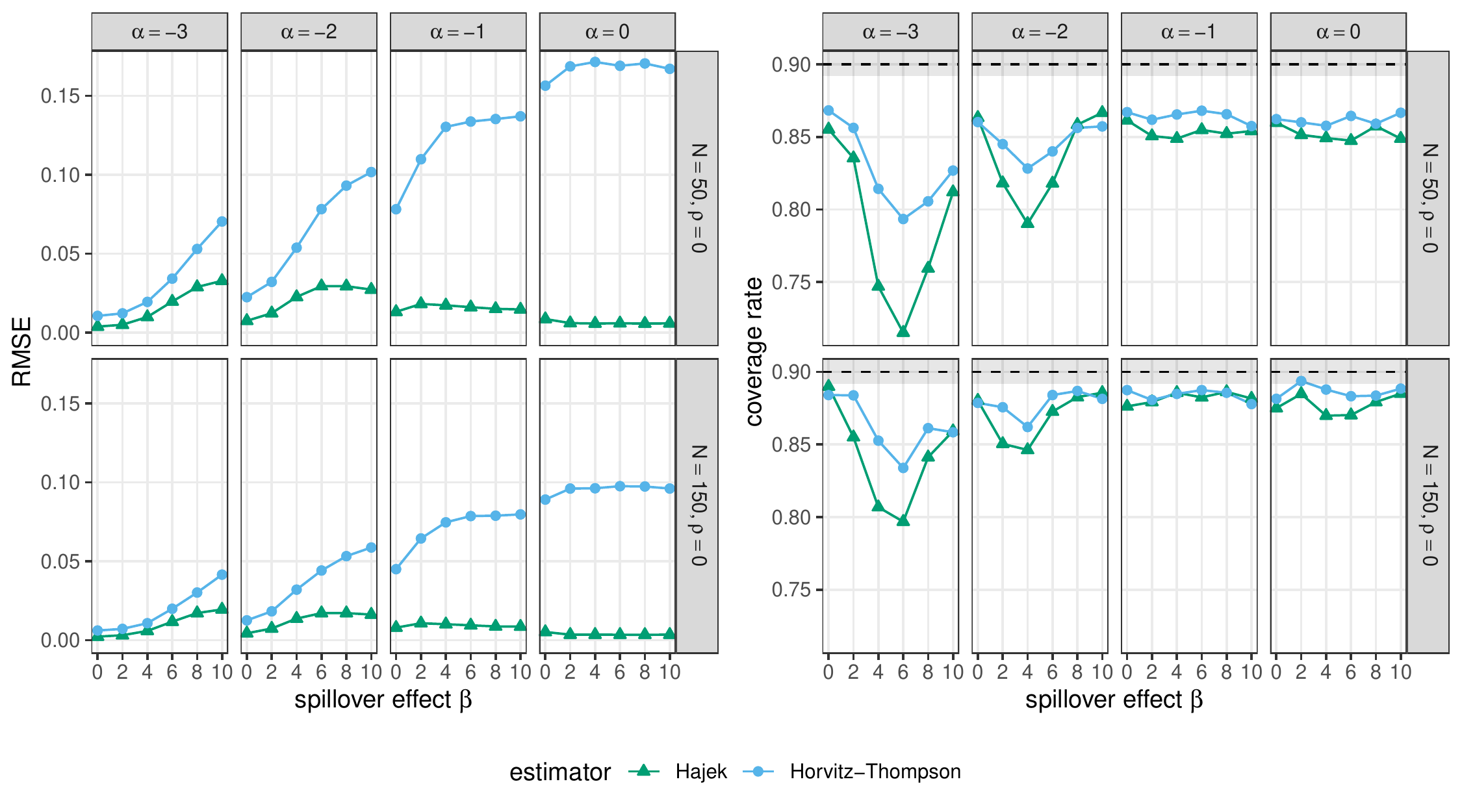}
\caption{(left) Root mean-squared-error of estimators in the off-policy simulation setup described in Section~\ref{sec:simulations-off-policy} (i.e., all villages assigned to random seeding). Columns vary the intercept $\alpha$, while rows vary the number of villages sampled. The horizontal axis varies the spillover effect $\beta$.
Error increases with $\beta$ as it produces more within-village dependence.
(right) Coverage rates for a 90\% nominal confidence interval; shaded area is the 95\% acceptance region ($p > 0.05$) for coverage being at least the nominal rate.
The confidence intervals have lower than nominal coverage, particularly for small sample sizes and rare outcomes.
}
\label{fig:sim-op-rmse-coverage}
\end{figure}

Figure \ref{fig:sim-op-rmse-coverage} (left) shows the superior performance of the H\'ajek estimator in RMSE. Note that compared with results in Figure \ref{fig:sim-rmse-coverage}, the H\'ajek estimator using only data from random seeding often has lower RMSE than the difference-in-means estimator using data from an experiment designed for this purpose.
In this off-policy simulation, the analytical asymptotic confidence intervals do not cover the true treatment effects at the advertised rate (Figure \ref{fig:sim-op-rmse-coverage}, left); this undercoverage is particularly evident for smaller sample sizes and rare outcomes. The undercoverage of analytical confidence intervals could motivate analysts to make use of exact finite sample inference as described in Section~\ref{sec:fish}. Note that the coverage of the H\'ajek confidence interval, where the underlying variance is itself only asymptotically correct, is generally worse.
 These results are consistent with the off-policy designs resulting in smaller effective sample sizes, a point we return to in the next subsection. 
 
\subsection{Design and effective sample size}
\label{sec:design_and_ess}

\begin{table}[t]
\centering
\begin{tabular}{c|c|ccc}
  \toprule
  & & \multicolumn{3}{|c}{Population effective sample size $\neffpop$}  \\
  Dataset & $N$ & $\tilde \tau$, Bernoulli$(0.5)$ & $\tilde \tau$, optimized & $\tilde \tau$, off-policy \\
  \midrule
Cai et al. & 150 & 625.8 & 865.8 & 385.5 \\ 
  AddHealth & 85 & 319.5 & 448.8 & 213.3 \\ 
  Banerjee et al. & 75 & 214.3 & 274.1 & 109.5 \\ 
  Paluck et al. & 56 & 351.6 & 539.0 & 299.9 \\ 
  Chami et al. & 17 & 37.9 & 47.9 & 10.8 \\ 
  \bottomrule
\end{tabular}
%OLD NUMBERS
% \begin{tabular}{c|c|cccc}
%   \toprule
%   & & \multicolumn{4}{|c}{Population effective sample size $\neffpop$}  \\
%   Dataset & $N$ & $\hat \tau_{DM}$, Bernoulli$(0.5)$ & $\tilde \tau$, Bernoulli$(0.5)$ & $\tilde \tau$, optimal & $\tilde \tau$, off-policy \\
%   \midrule
%   Cai et al. &  150 & 37.50 & 157.93 & 217.79 &  58.34  \\
%   \paluck{Paluck et al. &  56 & 14.00 & 87.90 & 134.76 &  32.06  \\}
%   \midrule
%   AddHealth &  85 &  21.25 & 79.87 & 112.20 & 32.76 \\
%   Banerjee et al. &  75 & 18.75 & 53.58 & 68.52 &  20.06 \\
%   Chami et al. & 17 & 4.25 & 9.46 & 11.98 & 2.33 \\
%   \bottomrule
% \end{tabular}
\caption{The population effective sample size $\neffpop$, calculated using 
equation~\eqref{eqn:neff-two-sample-population} from Appendix~\ref{app:ess}, for the H\'ajek estimator $\tilde \tau$ of the average treatment effect $\tausp$ (between random and one-hop targeting) on five collections of networks under different designs all targeting $k=2$ seeds.  The off-policy evaluation is for estimating one-hop targeting from random seeding data only. These $\neff^*$ can be interpreted as the number of villages needed for a difference-in-means estimator in a Bernoulli$(0.5)$ experiment to have the same precision.  The H\'ajek estimator always increases the effective sample size (in expectatation) over difference-in-means in a Bernoulli design, sometimes drastically.  For all datasets except \citet{chami2017social}, the H\'ajek estimator under an off-policy design has greater power than the difference-in-means estimator when an experiment is explicitly designed for the purpose of comparing strategies.}
\label{table:ess}
\end{table}

\begin{figure}
\centering
\includegraphics[width=9cm]{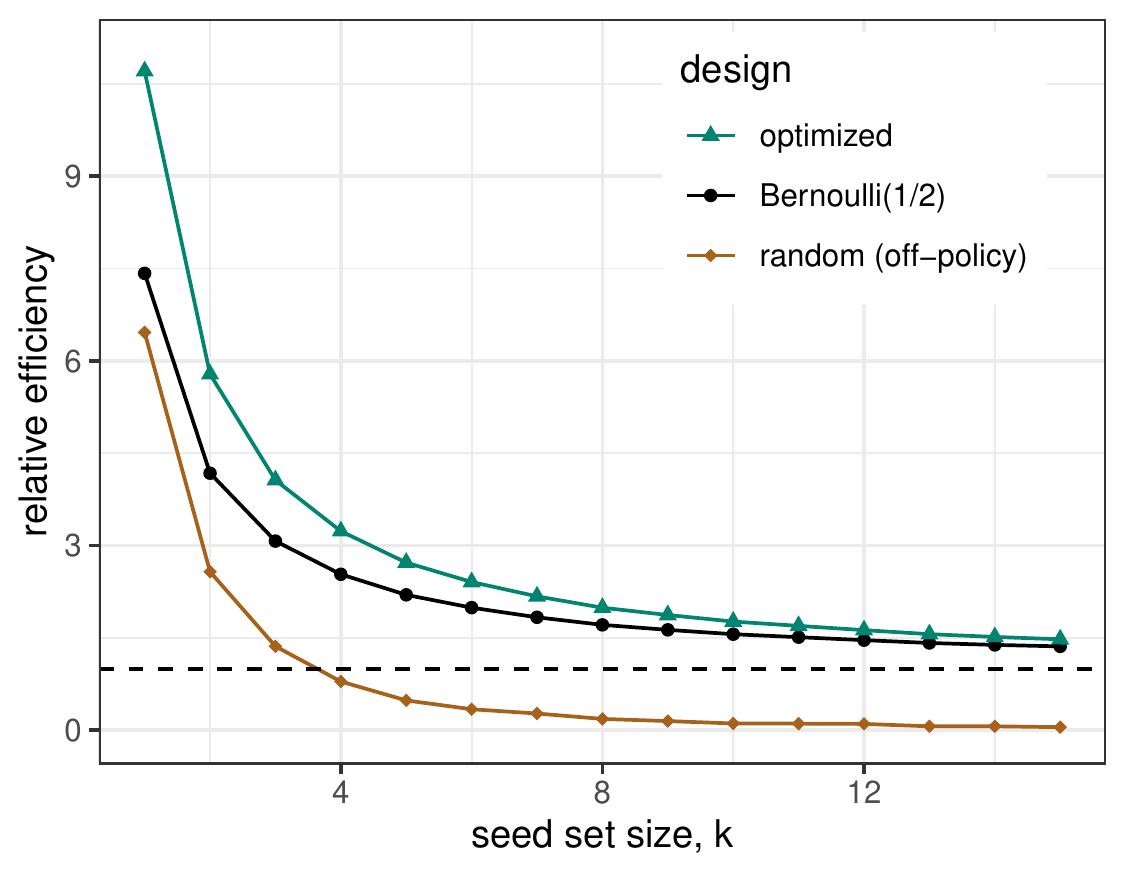}
\caption{The relative efficiency, defined as $\neffpop / N$, for comparing one-hop and random targeting under various designs for the~\citet{cai2015social} dataset of 150 villages, as a function of the seed set sizes $k$.  The points for $k = 2$ correspond to the values for~\citet{cai2015social} in Table~\ref{table:ess}.  The horizontal dashed line represents the nominal sample size $N$ and corresponds to a relative efficiency of 1.  For all points we record the efficiency gains of the H\'ajek estimator, equation~\eqref{eqn:hajek}, over using the difference-in-means estimator in a Bernoulli(0.5) experiment where, on average, 75 villages are assigned to each strategy, and vary only the seed set size $k$ and the design. We see major efficiency gains especially for small $k$.  Using the optimal design yields even more gains; in the extreme case, for $k=1$ the optimal design gives a more than 10-fold increase in effective sample size. 
As $k$ increases (for $k \ll n$) the one-hop distribution concentrates while the uniform distribution is spread thin across a combinatorially large support, a major cause of the drop in relative efficiency here.
Even estimation of the ATE under the off-policy design, in which all villages are assigned to random targeting, yields efficiency gains for small values of $k$.}
\label{fig:ess-k}
\end{figure}

We now examine how the effective sample size of the H\'ajek estimator varies as the design, network structure, and seed set size are varied. Our diagnostics are based on a population effective sample size, $\neffpop$, developed in detail in Appendix~\ref{app:ess}. In addition to the \citet{cai2015social} collection of networks, we analyze effective sample size for four additional collections of networks: the social networks of middle school students in New Jersey \citep{paluck2016changing}, the social networks of students in the AddHealth study in the United States \citep{resnick1997protecting}, the social networks of villages from a study of the diffusion of microfinance in India \citep{banerjee2013diffusion}, and the friendship networks of villages from a study of community health in Uganda \citep{chami2017social}. Calculating $\neff^*$ requires only knowledge of the network structure and no actual experimental outcome data; however, computing it requires assumptions about the outcomes, and we impose conditions on the outcomes that are most readily satisfied under a sharp null hypothesis (equation \eqref{eqn:sharp_null}). It is a useful single-number summary of precision gains or losses due to design and estimation choices. 

For each of these collections we first compute the population effective sample size $\neffpop$ for the H\'ajek estimator, as defined by equation~\eqref{eqn:neff-two-sample-population} in Appendix~\ref{app:ess}, with seed sets of size $k = 2$ under three different designs: a Bernoulli design, an optimized design, and an off-policy evaluation from only data assigned to random targeting. The optimized design, developed in detail in Appendix~\ref{app:opt-design}, minimizes the variance of of the H\'ajek  estimator under the sharp null hypothesis according to which the seed set does not affect outcomes. This assumption is strong, and thus we refer to the design as {\it optimized}, rather than optimal. The off-policy efficiency calculation differs slightly from the standard efficiency calculation used in importance sampling or survey science, because the target estimand here is an average treatment effect rather than a population mean.  Further detail is provided in Appendix~\ref{app:ess-op-ate}.  As a second evaluation, we focus on the \citet{cai2015social} collection of networks and extend our analysis beyond $k=2$ to larger seed set sizes. 

\subsubsection*{Results}

The results are given in Table~\ref{table:ess} and Figure \ref{fig:ess-k}. To interpret these effective sample sizes, by comparison a difference-in-means estimator $\hat{\tau}_{DM}$ for $N$ villages with a Bernoulli(0.5) design would have an effective sample size exactly equal to $N$. 

Table~\ref{table:ess} shows that using the H\'ajek estimator brings substantial increases in precision.  For example, for the \citet{cai2015social} dataset, using the H\'ajek estimator is equivalent to having run an experiment on 631 villages rather than the original 150, a fourfold increase.  Precision can be boosted further by using the optimized design described in Appendix~\ref{app:opt-design}.  Note also that for all but one of the data sets, the effective sample size for off-policy estimation is greater than that for na\"ive estimation with the Bernoulli design; that is, the proposed estimators yield greater precision from an experiment not designed for the purpose of comparing one-hop and random targeting ($\tilde \tau$, off-policy from random targeting) than a difference-in-means estimator for a field experiment conducted for that purpose ($\hat \tau_{DM}$, Bernoulli(0.5)).

The calculations in Table~\ref{table:ess} fix the seed set size at $k=2$. In Figure~\ref{fig:ess-k} we study how the size of the seed sets impacts the relative efficiency, defined as $\neff / N$, the ratio of the effective sample size to the nominal sample size.  We calculate the relative efficiency for $k=1,\ldots,15$ using probabilities calculated using the dynamic program discussed in Appendix~\ref{app:probs} and 1000 Monte Carlo sampled seed sets for each village. The relative efficiency is also equal to the inverse of the design effect~\citep{kish1965survey}.  
We observe that as seed sets become larger the selection probabilities under one-hop and random targeting quickly diverge on these empirical graphs with highly skewed degree sequences, reducing the benefits of both our H\'ajek estimator and optimized design. This reduction in power as $k$ grows suggests that designs involving smaller seed sets are better for testing hypotheses about differences between one-hop and random targeting; seeding with a small $k$ is also typically how these problems are posed.
That said, these effective sample size calculations are conducted under the null hypothesis and so do not take into consideration the possible social influence processes that may underly an outcome \citep{aral2013engineering}. For example, processes involving complex contagion~\citep{centola2007complex} should see very different differences between seeding strategy outcomes when evaluated using seed sets of size $k=1$ vs.\ sets of size $k>1$.

\section{Empirical Application\paluck{s}} 
\label{sec:empirical}
The proposed estimators can be applied to existing field experiments. First, they can be used to increase the precision of estimation in experiments that do directly compare two stochastic seeding strategies \citep{kim2015social}. Second, they can be used for off-policy estimation of contrasts between stochastic seeding strategies, even when, e.g.,~only random targeting was conducted, as is the case in the two field experiments we consider here.

\subsection{\citet{cai2015social}}
We use our method to provide a new analysis of the data studied in~\citet{cai2015social} in which we compare one-hop and random seeding.  The authors conducted a field experiment in villages in rural China to study peer effects in adoption of farmer's insurance.  Villagers were assigned to one of four groups that varied the timing and intensiveness of the marketing intervention, and the presentation of information about village-wide takeup in the case of later sessions. We take the seed set to be the set of villagers assigned to the ``intensive'' session at the first period.\footnote{\citet{cai2015social} describes the experiment was stratified on median household size and rice area. Exploratory analyses seem inconsistent with the most natural interpretation of this description. Thus, for now and for simplicity, we analyze the experiment as if the design were an unstratified, completely randomized experiment.
}
The response variable is the proportion of villagers who chose to purchase the farmer's insurance product.

There are a small number of edges between residents of different villages; we drop these edges and consider the villages to be entirely disjoint.  Not all villages had households assigned to treatments that varied within the village, and a few villages had insufficient network information; we drop all villages containing fewer than 25 edges.  After this preprocessing, we are left with 150 villages, which contrasts with the 185 villages originally analyzed by~\citet{cai2015social}.  Summary statistics for these 150 villages are given in Table~\ref{table:cai-summary-stats}.

\begin{table}[tb]
\centering
\begin{tabular}{lrrrr}
\hline
& mean & st.\ dev.\ & min & max \\ 
  \hline
Edges & 93.0 & 37.4 & 17.0 & 188.0 \\ 
  Nodes $n_i$ & 27.6 & 9.4 & 8.0 & 49.0 \\ 
  Mean in-degree & 3.4 & 0.7 & 1.2 & 4.7 \\ 
  St. dev. in-degree & 2.4 & 0.6 & 0.9 & 3.7 \\ 
  Treated \% & 22.9 & 7.0 & 10.0 & 50.0 \\ 
  Treated count $k_i$ & 6.2 & 2.3 & 1.0 & 13.0 \\ 
  $\log_{10} {n_i \choose k_i}$ & 12.7 & 4.9 & 2.3 & 26.3 \\ 
  Takeup \% & 44.9 & 19.5 & 5.1 & 95.8 \\ 
   \hline
  \end{tabular}
\caption{Summary statistics for the 150 villages from~\citet{cai2015social} analyzed here.}
\label{table:cai-summary-stats}
\end{table}

For each village, we compute the random and one-hop targeting probabilities of the observed seed set, conditional on the observed seed set size.  The random targeting probability is uniform across all seed sets of the same size, as in equation~\eqref{eqn:rand-prob}.  For the one-hop targeting probability in equation~\eqref{eqn:nom-prob2}, all of which involve large seed sets, we compute the normalizing probabilities via a dynamic program (as discussed briefly in Section~\ref{sec:computingprobs} and more extensively in Appendix~\ref{app:probs}).
Many observed seed sets (67 of 150)  
are not possible under one-hop targeting because they include nodes with zero in-degree. Aside from these cases the order of magnitude of the probabilities for both strategies are mostly determined by the size of the seed set, but there is enough discrepancy between the probabilities to facilitate off-policy estimation.

\begin{figure}[tb]
\centering
\includegraphics[height=8cm]{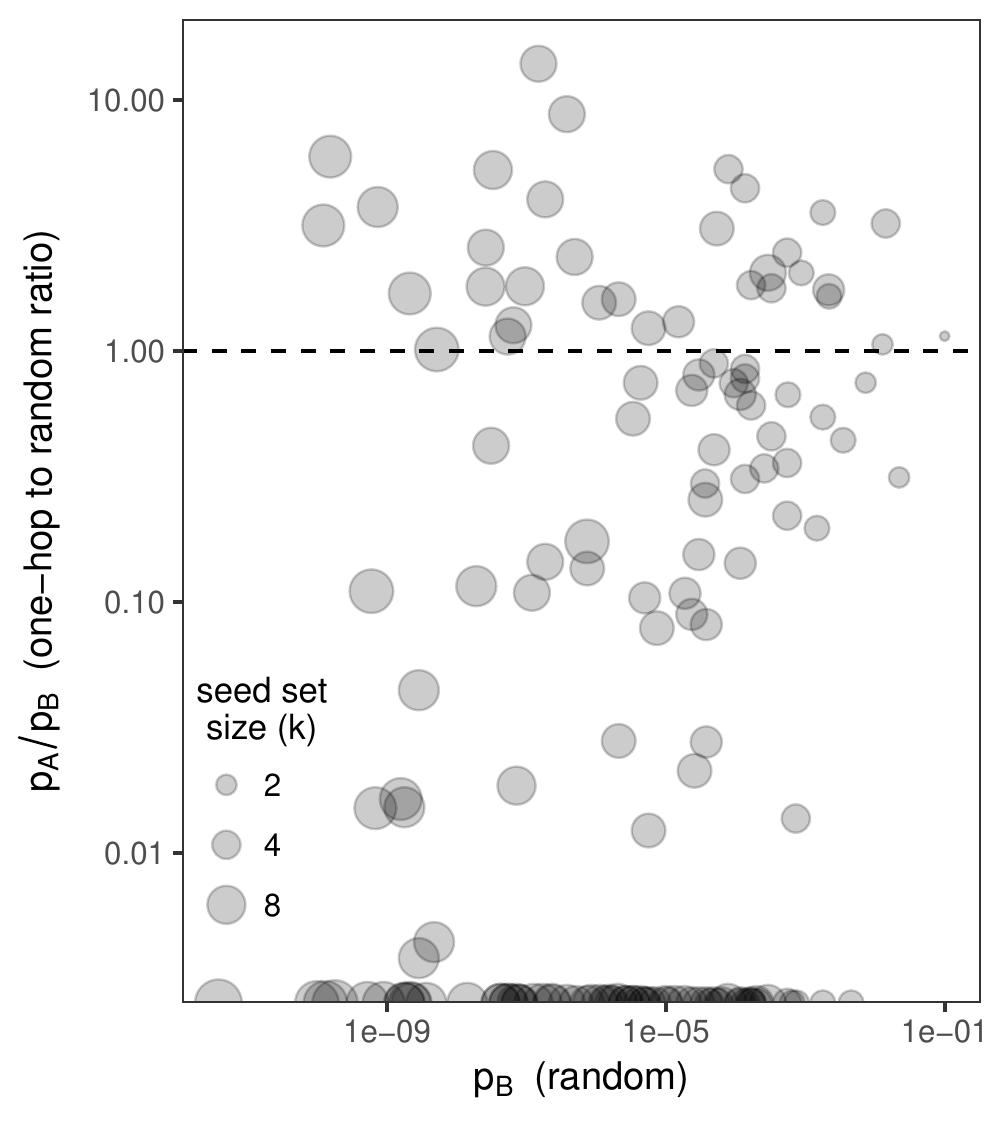}
\includegraphics[height=8cm]{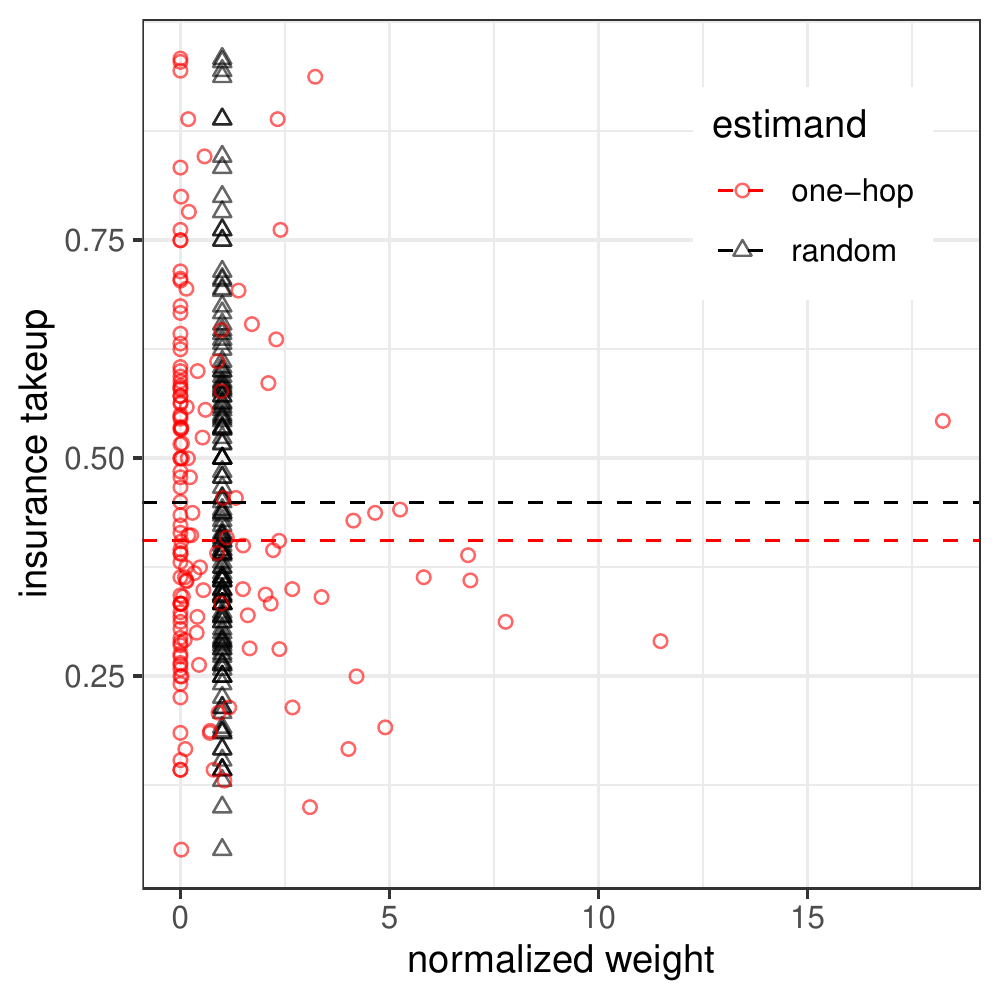}
\caption{(left) The ratio of one-hop and random targeting probabilities for the 150 villages analyzed from the~\citet{cai2015social} study.  Absolute probabilities are correlated with seed set size, but there is considerable variation in the ratio. Seed sets with $p_B=0$ are plotted at the bottom of the vertical axis. (right) The outcome (fraction of households buying insurance) as a function of the normalized (H\'ajek) weights. Since the seed sets in the study were assigned via random targeting, the estimate for that strategy is an unweighted sample mean, whereas the one-hop targeting estimate applies reweighting. The horizontal dashed lines are the (H\'ajek) estimated mean outcomes under each strategy.}
\label{fig:cai-plot}
\end{figure}

We compute the weights of the importance sampling estimators, $w_i^A$ and $w_i^B$, from equation~\eqref{eqn:weights}.  Since the seed sets were assigned according to random targeting, the random targeting weights are constant, $w_i^B = b_i / b_i = 1$.  The one-hop targeting weights are the ratio of one-hop to random targeting probabilities, $w_i^B = a_i / b_i$.
Figure~\ref{fig:cai-plot} displays the probabilities and their ratio (left) and the normalized weights ($\tilde{w}_i^A, \tilde{w}_i^B$) used by the H\'ajek estimator (right).

Table~\ref{table:cai-results} shows the H\'ajek estimate and associated inference. Asymptotic (analytic) inference would lead to the conclusion that the one-hop strategy would \emph{reduce} takeup of insurance by 0.3 to 8.5 percentage points. Bootstrap inference leads to more conservative but still suggestive conclusions. Given the simulation results in Section \ref{sec:simulations}, in which we observed undercoverage in some settings for off-policy estimation, we should be cautious in relying on this statistical inference without further analysis.
First, the one-hop targeting estimator has an effective sample size of $\neff = 28.5$ using the off-policy effective sample size expression given in equation~\eqref{eqn:neff-one-sample}. The random targeting estimator of course has $\neff = n = 150$. 

This small effective sample size for the one-hop targeting estimator suggests a great deal of caution in using the estimated variance to conduct inference (e.g., to construct confidence intervals) based on normal theory as we have done here.
Thus, we also conduct a hypothesis test using Fisherian randomization inference as discussed in Section~\ref{sec:fish}. This test also provides some evidence against the null of no effects of choice of seeds.

\begin{table}[t]
\centering
\begin{tabular}{rr}
  \hline
estimate  (one-hop $-$ rand) & -0.0436 \\ 
  SE (analytic) & 0.0209 \\ 
  SE (bootstrap) & 0.0257 \\ 
  95\% CI (analytic) & [-0.0846, -0.0027] \\ 
  95\% CI (bootstrap) & [-0.0909, 0.0088] \\ 
  p-value (analytic) & 0.0367 \\ 
  p-value (Fisherian) & 0.0974 \\ 
   \hline
\end{tabular}
\caption{H\'ajek estimate and inference for the difference in insurance takeup rates between one-hop and random seeding for \citet{cai2015social}, which provide some evidence that one-hop seeding would have \emph{reduced} adoption of insurance.}
\label{table:cai-results}
\end{table}

This analysis both demonstrates how our proposed estimators can be used off-policy and provides some cautionary results compared with previous evidence about the one-hop targeting strategy~\citep{kim2015social}. In particular, these results suggest that one-hop seeding may in some relevant settings perform no better or even worse than simple random seeding.

\paluck{
\subsection{\citet{paluck2016changing}}
\label{sec:paluck}

\citet{paluck2016changing} conducted a field experiment in 56 middle schools in New Jersey, in which they randomly assigned an intervention designed to reduce bullying and other peer conflict. We analyze data from the 28 schools assigned to treatment. Summary statistics for these schools are given in Table \ref{table:paluck-summary-stats}. A within-school randomization then assigned some students to be seeds: these students were invited to participate in a program that encouraged them to take a public stance against conflict among their peers at school.  \citet{paluck2016changing} measure several outcome variables of interest; here we focus on the number of peer conflict events per student as measured by administrative reports, and defer the results for other outcomes to Appendix~\ref{app:paluck}.  For peer conflict events, lower values of the outcome are viewed as desirable.

\begin{table}[t]
\centering
\begin{tabular}{lrrrr}
\hline
& mean & st.\ dev.\ & min & max \\ 
  \hline
  Edges & 3157.5 & 1395.3 & 1046.0 & 6561.0 \\ 
  Nodes $n_i$ & 426.4 & 172.9 & 138.0 & 835.0 \\ 
  Mean in-degree & 7.3 & 0.5 & 5.7 & 8.1 \\ 
  St. dev. in-degree & 4.4 & 0.4 & 3.7 & 5.2 \\ 
  Treated \% & 6.8 & 2.3 & 3.8 & 14.5 \\ 
  Treated count $k_i$ & 26.0 & 4.6 & 20.0 & 32.0 \\ 
  $\log_{10} {n_i \choose k_i}$ & 94.5 & 23.1 & 54.8 & 133.1 \\ 
  Peer conflict rate ($\times 100$) & 15.2 & 13.4 & 0.0 & 49.1 \\ 
   \hline
  \end{tabular}
\caption{Summary statistics for the 28 treatment schools from~\citet{paluck2016changing} analyzed here.}
\label{table:paluck-summary-stats}
\end{table}

\begin{figure}[tb]
\centering
\includegraphics[height=7cm]{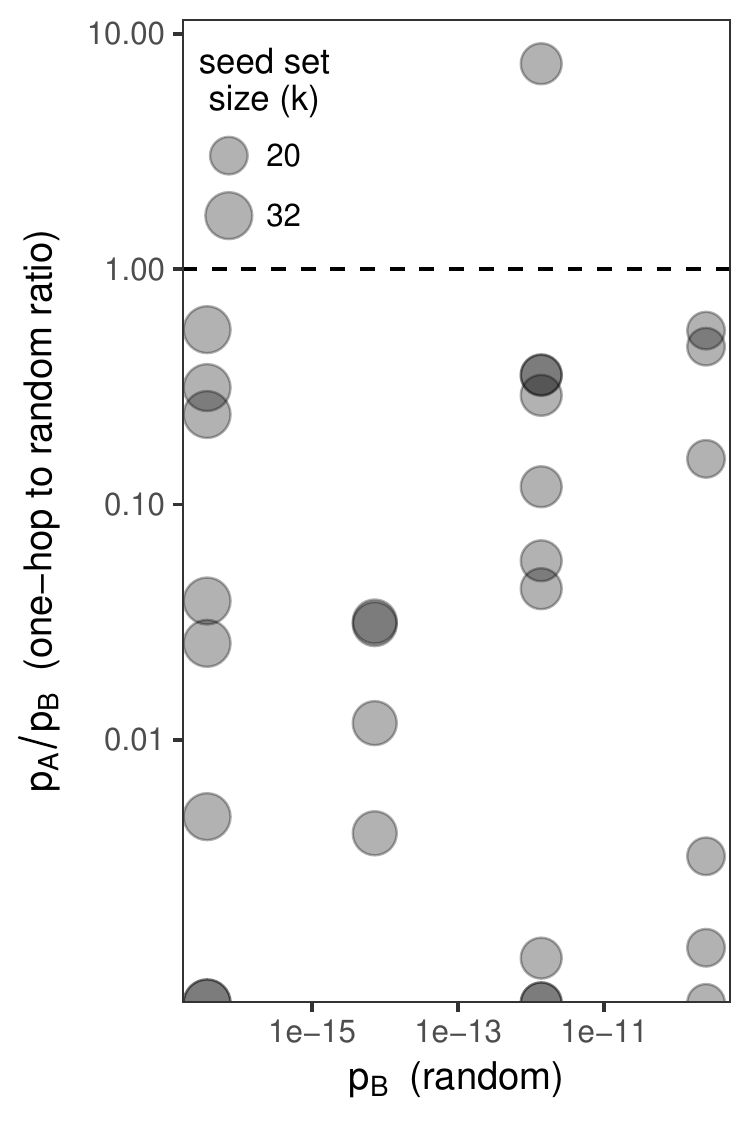}
\includegraphics[height=7cm]{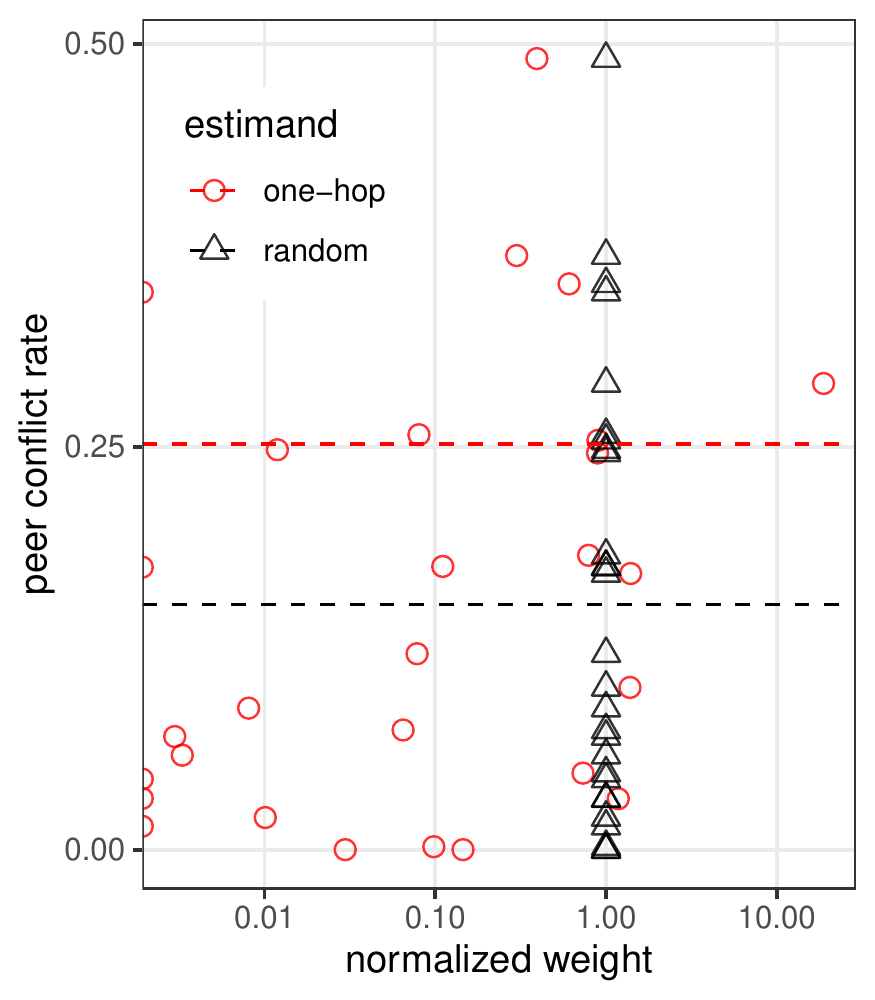}
\includegraphics[height=7cm]{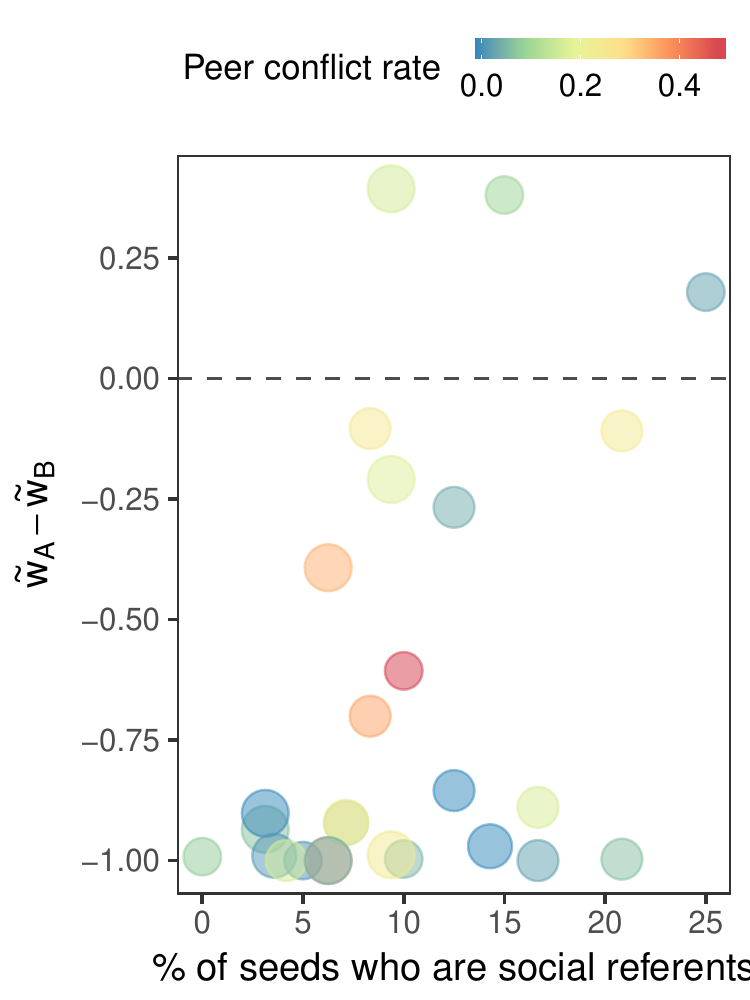}
\caption{(left) The ratio of one-hop and random targeting probabilities for the 28 treated schools from the~\citet{paluck2016changing} study.  One school has a seed set with higher probability under one-hop than random seeding.
(center) The primary outcome, number of administrative peer conflict reports per student, as a function of the normalized weights. The vertical dashed lines are the (H\'ajek) estimated means.
(right) Relationship between measure of seed centrality used by \citet{paluck2016changing} and the difference in weights used by our estimator. One school with very large positive $\tilde{w}_i^A - \tilde{w}_i^B$ (17.8) is not shown; 17\% of its seeds were social referents.
}
\label{fig:paluck-plot}
\end{figure}

Like our analysis of \citet{cai2015social}, this analysis is also an off-policy evaluation, but \citet{paluck2016changing} treat a smaller fraction of nodes, making it perhaps more typical of seeding with a limited budget. This intervention was also hypothesized to be more effective if more central individuals were seeds. In fact, \citet{paluck2016changing} find that treatment reduces peer conflict more when a larger fraction of the seeds that are ``social referents,'' defined as being in the top decile of in-degree for that school. Thus, one might expect that one-hop seeding would be effective in this setting.

\citet{paluck2016changing} use a blocked (stratified) randomization to balance selected seed sets by four blocks formed by grade and gender. We thus consider a variation on one-hop seeding that conditions on selecting the observed number of seeds $k_{ib}$ in each block $b$ of school $i$; we could think of this as reflecting a desire to ensure there are seeds in each grade and gender.
As with the previous analysis, Figure \ref{fig:paluck-plot} (left) shows the ratio of probabilities $p_A / p_B$ of the observed seed sets, illustrating that some (5 of 28) schools have seed sets that are impossible to reach under one-hop seeding. There is important heterogeneity in how probable the observed seed sets are under each strategy, though only one observed seed set is more probable under one-hop than random targeting. 

\begin{table}[t]
\centering
\begin{tabular}{rr}
  \hline
estimate  (one-hop $-$ rand) & 0.0997 \\ 
  SE (analytic) & 0.0231 \\ 
  SE (bootstrap) & 0.0432 \\ 
  95\% CI (analytic) & [0.0543, 0.1451] \\ 
  95\% CI (bootstrap) & [0.0098, 0.1542] \\ 
  p-value (analytic) & 1.7e-05 \\ 
  p-value (Fisherian) & 0.0846 \\ 
   \hline
\end{tabular}
\caption{H\'ajek estimate and inference for the difference in peer conflict per student one-hop and random seeding for \citet{paluck2016changing}, which provide some evidence that one-hop seeding would have \emph{increased} peer conflict (i.e., an undesirable outcome).}
\label{table:paluck-results}
\end{table}

Table~\ref{table:paluck-results} shows the H\'ajek estimate of the average treatment effect on peer conflict of one-hop vs.\ random targeting and associated inference. Asymptotic and bootstrap inference would lead to the conclusion that the one-hop strategy would \emph{increase} rates of peer conflict, as measured by administrative reports, by 0.01 to 0.15 incidents per student. Given the small number of schools, we conduct Fisherian randomization inference, which also provides some evidence against the null of no effect of choice of seed set. These results again suggest that one-hop seeding may in some cases perform worse than simple random seeding.

Appendix \ref{app:paluck} includes results for other secondary outcomes, such as self-reported wearing of wristbands that indicate a commitment to reducing conflict. These results, like those in \citet{paluck2016changing} for the fraction of social referents, do not provide much evidence against the null, at least when using Fisherian randomization inference, but have a sign more favorable to one-hop seeding.
}

\begin{comment}
\subsection{\citet{kim2015social}}
We expect that future versions of this paper will include reanalysis of the data from \citet{kim2015social}, illustrating ``on-policy'' evaluation as that study did compare random and one-hop targeting.
\end{comment}

\subsection{Implications for one-hop seeding}

One-hop seeding, like other (mostly deterministic) strategies, is typically motivated by appeal to theory or simulations with an assumed model of social contagion (e.g., our own simulations shown in Figure~\ref{fig:sim-true-tau}). In the above analyses, the empirical performance of one-hop seeding is less clear than theory tends to predict.

The basic theoretical premise of one-hop targeting is that by seeding an intervention at nodes with higher than average in-degrees, more nodes in the network will have a direct social connection to the intervention and hopefully adopt themselves. The main appeal of one-hop targeting, towards this goal, is its feasibility without complete knowledge of the network; if one had access to a full social network survey at the onset of the intervention then one could choose to directly target the highest in-degree individuals. But besides being conveniently feasible without a full social network survey, one-step targeting has a specific potential advantage over targeting the maximum in-degree individuals: the maximum in-degree individuals are often tightly clustered in the network---as is common in networks with core-periphery structure \citep{borgatti2000models,rombach2017core}---while the high in-degree individuals selected by one-hop targeting will be relatively spread out due to the initially random ``nominators.'' Thus, the neighbors of a seed set selected by one-hop targeting is likely to be less redundant and perhaps more influential as a whole\footnote{Avoiding redundancy is key to influence maximization in the widely studied independent cascade and linear threshold models~\citep{kempe2003maximizing}. However, if individual decision processes resemble a complex contagion process~\citep{centola2007complex} then such spreading of seeds would in fact be less desirable than explicitly targeting high degree individuals.}
\citep[cf.][]{kim2015social}, relative to directly targeting the highest in-degree individuals. 

If widely accepted theories motivate why one-hop targeting should lead to a higher adoption rate than random targeting (for interventions seeking to maximize adoption),  why do our results suggest that one-hop targeting is no more effective (and possibly less effective) than random targeting? Here we offer are a number of possible explanations, none of which are definitive and all of which suggest important follow-up work. 

First, it is possible that the social networks collected from the surveys in these studies are not the networks that matter in terms of influence processes guiding the relevant adoption decisions. The study of name generators \citep{campbell1991name,perkins2015social} in sociology has long established that different questions lead to different networks, e.g.~``Who are your friends?''\ vs.~``With whom do you discuss important matters?''\ \citep{bearman2004cloning}. Some name generators have a tendency to elicit strong ties while others elicit weak ties~\citep{momeni2017inferring}. It is well known that strong and weak ties figure differently in information diffusion and social decision making~\citep{rapoport1961study,granovetter1973strength,mcadam1986recruitment}. If trying to maximize adoption, it is natural to then ask what name generator leads to the greatest adoption under one-hop targeting, and also quite natural that one-hop targeting paired with some name generators would lead to less adoption than random seeding. In this vein, \citet{chami2017social} asked both about close friends and about trusted sources of health advice. \citet{banerjee2013diffusion} collected responses to a total of twelve different name generators, although most analyses of that study (including our use of these networks in Section~\ref{sec:design_and_ess}) analyze only the flattened network of ``all relationships.''

Questions about name generators raise an important dimension in which one-hop targeting can be modified to ask more ``ambitious'' questions when it is actually being deployed as a seeding strategy in the absence of an existing network survey. One could instead ask subjects to name ``the most respected farmers'' (for the weather insurance experiment) or ``the students most people look up to'' (for the anti-bullying intervention). This strategy would still count as a stochastic seeding strategy (when the surveyed people are a random set of individuals), and many of the ideas in this paper could be applied to evaluate such strategies. As described in the introduction, \citet{banerjee2017using} study a variant of one-hop; there they use this kind of ambitious question (``If we want to spread information about a new loan product to everyone in your village, to whom do you suggest we speak?''). This can be regarded as an explicitly stochastic variation on name generators used for identifying ``opinion leaders'' \citep{flodgren2011local}.

As a second possible explanation, even if the seeds selected by one-hop targeting may be more influential, they may not be susceptible to the initial intervention. \citet{aral2012identifying} present some evidence of (negative) correlations between susceptibility and influence; this can have substantial consequences for approximately optimal seeding \citep{aral2018social}. Similarly, \citet{bakshy2011everyone} highlight that if more influential individuals require larger inducements to adopt or promote a behavior, then targeting them may be a poor use of a limited budget.

As a third, more speculative possible explanation of our findings, it could be that the behavior in our interventions spread via a ``push'' mechanism (the seed needs to tell people about the intervention for it to spread), as opposed to a ``pull'' mechanism (the friends of the seed observe them). The sharply different dynamics of diffusion processes under push and pull mechanisms have been widely studied in the computer science literature~\citep{demers1987epidemic,chierichetti2011rumor}. A behavior that spreads via a push mechanism would benefit from being seeded at nodes with high out-degree, as opposed to high in-degree for a pull mechanism. While the in-degree and out-degree of nodes are often correlated, it is generically possible that a seeding strategy that climbs in average in-degree could decline in average out-degree. Again, this possibility highlights the importance of the specific name generator and survey technique used. In the \citet{paluck2016changing} surveys, out-degree was effectively capped as students were asked for up to ten students they chose to spend time with; over 40\% of students named exactly ten such students.

As a final variation on concerns about the surveyed network possibly being the ``wrong'' network, it is possible that the actual social networks describing influential relationships in both these settings are regular graphs (everyone has the same number of influential relations) or nearly regular. In a recent study of the friendship paradox and contact strength, \citet{bagrow2017friends} analyzed contact frequencies (as a proxy for tie strength) on Twitter and in cellular phone networks and found that networks of frequent ties are nearly regular, leading to a tempered friendship paradox: ``your closest friends have only slightly more friends than you do.'' A finding of little or no difference in adoption rates between random and one-step targeting would be consistent with the adoption decisions in the two field experiments we analyzed here relying on social networks that are nearly regular. Regular networks cannot, however, explain why one-hop targeting might be less effective than random targeting, and the \citet{cai2015social} networks had sufficient variation in in-degree to generate positive effects of one-hop in our simulations in Section \ref{sec:simulations}.

It is important to stress that our empirical findings around one-hop targeting are not inconsistent with established findings that one-hop targeting can be successfully leveraged to design efficient \emph{sensing} strategies \citep{christakis2010social, garcia2014using, kryvasheyeu2015performance}. In prior field work on epidemiological outbreak detection using ``one-hop measurement'' by \citet{christakis2010social}, instances of the flu occurred earlier in a population selected by a one-hop strategy than a random population. But, in the language of the present discussion, that finding only means that the one-hop strategy was successful in reaching a population that was more (epidemiologically) susceptible, but not necessarily (epidemiologically) influential.

\section{Conclusions}
\label{sec:discussion}
Stochastic seeding strategies are an appealing way of leveraging network structure for marketing and behavior change interventions in commerce, public health, education, economic development, and management when faced with limited existing network information and a limited budget. Besides the applications in public health marketing, marketing in the developing world, and behavior change in education considered at length here, this occurs in several settings of long-standing interest to researchers in marketing and related fields, such as decisions by doctors (e.g., to prescribe pharmaceuticals) \citep[e.g.,][]{coleman1957diffusion,iyengar2011opinion} and households installing solar panels \citep[e.g.,][]{bollinger2012peer,kraft2018credibility,bollinger2019visibility}; in both cases, the relevant networks may be hard to determine. For example, both interventions in and research on social contagion in doctors' decisions typically involved costly surveys of doctors. Even in setting where many relationships are articulated in online social networks, marketers typically have only limited access to these networks (e.g., crawling the Twitter follower network).

The present approach shows how this costly network information can be leveraged --- even while contemplating larger-scale interventions that cannot count on full network information in deployment.
In particular, many of these settings involve multiple markets or municipalities, or multiple parallel promotions within a market, which facilitate evaluating multiple seeding strategies. These seed sets could be assigned according to assigning different markets to different seeding strategies, or even via the kind of optimized design we developed here. Furthermore, if researchers use a design that places positive probability on many seed sets, it will be possible to evaluate seeding strategies that were not contemplated until after the experiment.

In our development of off-policy evaluation for seeding, we show that evaluations of stochastic strategies is possible with data arising from, e.g., unconditional random assignment. Beyond one-hop seeding, our off-policy evaluation framework opens the door to broad investigations of other stochastic seeding strategies that can be studied using data collected under random assignment.

On the matter of one-hop seeding, by analyzing the field experiments by \citet{cai2015social} and \citet{paluck2016changing} (both of which used random seeding), we find evidence in both cases that the efficacy of the interventions would have been unchanged or slightly reduced under one-hop seeding. 
In both cases,
characteristics of the setting and the original results suggested that one-hop seeding would be a promising way to increase the desired outcomes.
This emphasizes the importance of credible empirical evaluation of these strategies.

A much larger follow-up experiment to \citet{kim2015social}, registered in \citet{shakya2017exploiting}, employs the same basic design as its predecessor. Given our simulations here, we expect the preregistered design and analysis will have lower power than achievable through better design and analysis. That said, the analysis developed in this work is fully applicable as a post hoc analysis. %\jucom{Dean, agree with this?} \decom{Maybe we ought to pre-register it!} \jucom{Maybe an arxiv comment counts as pre-registration...}
Our hope is that our perspective on stochastic interventions will in turn inform the design and analysis of future experiments, as well as the practice of seeding in marketing, public health, education, and development economics.

\pagebreak
%\printbibliography
\bibliography{seedtargeting}

\pagebreak

\appendix

\counterwithin{figure}{section}
\counterwithin{table}{section}

\section{Proofs}
\label{app:proofs}
\subsection{Proof of Proposition~\ref{prop:ht}}

\begin{proof}
To show unbiasedness, it suffices consider a single seed set $S \sim p_\Delta$, since the seed sets are sampled iid.  For a single seed set $S$, the Horvitz--Thompson estimator~\eqref{eqn:ht} is 
\[\hat \tau = (W_A - W_B)Y = \frac{(p_A(S) - p_B(S))y(S)}{p_\Delta(S)}.\]This quantity has expectation
\[\E[\hat \tau] = \E_\Delta \left[\frac{p_A(S)y(S)}{p_\Delta(S)}\right] - \E_\Delta \left[\frac{p_B(S)y(S)}{p_\Delta(S)}\right] = \E_A[y(S)] - \E_B[y(S)] = \tausp.\]
The variance of $\hat \tau$, for a sample of size $N$, is calculated as
\begin{align*}
\var(\hat \tau) &= \var\left[\avgin (w_i^A - w_i^B)y_i\right] = \frac{1}{N} \E \left[\frac{(P_A - P_B)^2Y^2}{P_\Delta^2}\right] - \tausp^2 \\
&= \frac{1}{N} \E \left[\frac{1}{P_\Delta^2}\left((P_A - P_B)Y - \tausp P_\Delta\right)^2\right] = \frac{1}{N}\E[ ((W_A - W_B)Y - \tausp)^2] = V_{\hat \tau} / N,
\end{align*}
where $V_{\hat \tau}$ is defined in equation~\eqref{eqn:ht-variance}.
\end{proof}

\subsection{Proof of Proposition~\ref{prop:hajek}}

\begin{proof}
Denote the sample averages
\begin{align*}
\bar w_A &= \avgin w_i^A \\
\bar w_B &= \avgin w_i^B \\
\overline{wy}_A &= \avgin w_i^A y_i \\
\overline{wy}_B &= \avgin w_i^B y_i
\end{align*}
so that
\[\tilde \tau = \frac{\overline{wy}_A}{\bar w_A} - \frac{\overline{wy}_B}{\bar w_B}.\]
Consistency follows from the consistency of these sample averages and the continuous mapping theorem.

The normality result is a straightforward but slightly tedious application of the delta method, and follows the standard approach for characterizing the limiting behavior of ratio estimators (see, for example,~\citet[Section 5.6]{sarndal1992model} or~\citet[Ch. 9]{mcbook}).
Let $\hat \beta = (\bar w_A, \bar w_B, \overline{wy}_A, \overline{wy}_B)^\top$, and notice that $\beta := \E[\hat\beta] = (1, 1, \mu_A, \mu_B)$.  Furthermore $\sqrt{N}(\hat \beta - \beta) \wto \mathcal{N}(0, \Sigma)$, where the entries of the asymptotic variance $\Sigma$ are defined for $\Omega = A, B$, by
\begin{align*}
\sqrt{N}\var(\bar w_\Omega) &= \E\left[\frac{P_\Omega^2}{P_\Delta^2}\right] - 1 \\
\sqrt{N} \var(\overline{wy}_\Omega) &= \E \left[\frac{P_\Omega^2 Y^2}{P_\Delta^2}\right] - \mu_\Omega^2 \\
\sqrt{N} \cov(\bar w_\Omega, \overline{wy}_\Omega) &= \E \left[\frac{P_\Omega^2 Y}{P_\Delta^2}\right] - \mu_\Omega \\
\sqrt{N} \cov(\bar w_A, \bar w_B) &= \E \left[\frac{P_A P_B}{P_\Delta^2}\right] - 1 \\
\sqrt{N} \cov(\overline{wy}_A, \overline{wy}_B) &= \E \left[\frac{P_A P_B Y^2}{P_\Delta^2}\right] - \mu_A\mu_B \\
\sqrt{N} \cov(\bar w_A, \overline{wy}_B) &= \E \left[\frac{P_A P_B Y}{P_\Delta^2}\right] - \mu_B \\
\sqrt{N} \cov(\bar w_B, \overline{wy}_A) &= \E \left[\frac{P_B P_A Y}{P_\Delta^2}\right] - \mu_A
\end{align*}
where $S \sim P_\Delta$.
Now $\tilde \tau = f(\hat \beta)$, where $f(a, b, c, d) = c/a - d/b$ and the gradient evaluated at $\beta$ is $\nabla_\beta f = (-\mu_A, \mu_B, 1, -1)$.  Then by the delta method, 
\begin{align*}
V_{\tilde \tau} = \nabla_\beta f^\top \Sigma \nabla_\beta f 
=& \mu_A^2 \left(\E\left[\frac{P_A^2}{P_\Delta^2}\right] - 1\right) + \mu_B^2\left(\E\left[\frac{P_B^2}{P_\Delta^2}\right] - 1\right) + \E \left[\frac{P_A^2Y^2}{P_\Delta^2}\right] - \mu_A^2 + \E \left[\frac{P_B^2Y^2}{P_\Delta^2}\right] - \mu_B^2 \\
&- 2\mu_A \left(\E\left[\frac{P_A^2 Y}{P_\Delta^2}\right] - \mu_A\right) - 2\mu_B \left(\E\left[\frac{P_B^2 Y}{P_\Delta^2}\right] - \mu_B\right) \\
&- 2\mu_A\mu_B \left(\E\left[\frac{P_A P_B}{P_\Delta^2}\right] - 1\right) - 2\left(\E \left[\frac{P_A P_B Y^2}{P_\Delta^2}\right] - \mu_A\mu_B\right) \\
&+ 2\mu_A \left(\E \left[\frac{P_A P_B Y}{P_\Delta^2}\right] - \mu_B\right) - 2\mu_B \left(\E \left[\frac{P_B P_A Y}{P_\Delta^2}\right] - \mu_A\right). \\
=& \E \bigg[\frac{1}{P_\Delta^2} \bigg(\mu_A^2 P_A^2 + \mu_B^2 P_B^2 + 2\mu_A P_A Y(P_B - P_A) - 2 \mu_B P_B Y(P_A + P_B)\\
&+ P_A^2 Y^2 + P_B^2Y^2 - 2P_A P_B Y^2 - 2 \mu_A \mu_B P_A P_B\bigg)\bigg].
\end{align*}
Rearranging terms produces the expression in equation~\eqref{eqn:hajek-variance}.
\end{proof}

\newpage
\section{Computing seed set probabilities}
\label{app:probs}

\subsection{A dynamic program for computing one-hop targeting}

As discussed in Section~\ref{sec:computingprobs}, the seed set probabilities under one-hop seeding strategy with replacement (of seeds) are given by
\begin{equation}
\label{eqn:nom-proba}
\P_i^{A,\text{repl}}(S_i = s) = k! \prod_{v \in s} \frac{1}{n_i} \sum_{u \in \mathcal N_{\text{in}} (v)} \frac{1}{d_u^{\text{out}}},
\end{equation}
where $\P_i^{A,\text{repl}}$ denotes probability with respect to $p_i^A$ with replacement, $\mathcal N_{\text{in}} (v)$ denotes the set of in-neighbors of $v$, and $d_u^{\text{out}}$ denotes the out-degree of node $u$.

Recall that to translate probabilities with replacement to probabilities without, we need to compute the total probability 
\begin{equation}
\label{eqn:nom-suma}
\pi_i = \sum_{s \in \mathcal S_i : s \text{ unique}, |s|=k} \P_i^{A,\text{repl}}(S_i = s),
\end{equation}
which lets us use a simple normalization to compute the one-hop targeting probabilities without replacement:
\begin{equation}
\label{eqn:nom-prob2a}
\P_i^A(S_i = s) = \frac{1}{\pi_i} \P_i^{A,\text{repl}}(S_i = s) .
\end{equation}

Here we discuss a dynamic program for computing $\pi_i$, which is suitable when $|\mathcal S_i| = \binom{n_i}{k}$, the number of seed sets of size $k$ (the number of terms in the sum \eqref{eqn:nom-suma} to compute $\pi_i$), is large. More specifically, we give a dynamic programming approach to exactly compute $\pi_i$ in time $O(n_i k + m_i)$, where $m_i=|E_i|$ is the size of the edge set of village $i$. Contrast this runtime with the runtime of a naive summation, which is exponential in $k$. We emphasize, as noted in the main text, that this dynamic programming solution can be adapted to many network-aware stochastic policies that decompose cleanly into node-level targeting probabilities that are themselves tractable.

Define 
\[
f_v = \sum_{u \in \mathcal{N}_{in}(v)} \frac{1}{d_u ^{out}},
\]
which we assume to have been computed already (a step that takes $O(m_i)$ time). Then we have
\begin{equation}   \label{Eq:der-pi_i}
\pi_i = \frac{k!}{n_i} \sum_{s \subseteq V: \lvert s \rvert = k}~ \prod_{v \in s} f_v
\end{equation}
Now let $A = [f_1, f_2, \dots, f_{n_i}]$ be an arbitrary ordering of $\{f_v\}_{v \in V}$, 
and denote $A_j = [f_1, \dots, f_j]$ as the vector consisting of the leading $j$ elements in $A$. Notice that $A = A_{n_i}$.

To construct our dynamic programming solution, we define a subproblem
\begin{equation}   \label{Eq:def-sub_problem}
S(j, \ell) = \sum_{s \subseteq A_j: \lvert s \rvert = \ell}~ \prod_{v \in s} f_v,
\end{equation}
i.e., the probability that if there were only $j$ elements in $A$ and we sought to compute the sum of the product of every subset of size $\ell$. Notice then that $\pi_i = \frac{k!}{n_i} \cdot S(n_i, k)$.

Now we give a recursive formula for $S(j,\ell)$. In searching all subsets with $\ell$ elements of $A_j = [f_1, \dots, f_{j-1}, f_j]$, element $f_j$ may or may not be included in the subsets that we sum across. There are simply two scenarios we need to consider:
\begin{itemize}
\item If $f_j$ is included, we will choose $\ell-1$ elements from $A_{j-1} = [f_1, \dots, f_{j-1}]$, and thus the total sum is $S(j-1, \ell-1) \cdot f_j$.
\item If $f_j$ is not included, then we choose $\ell$ elements from $A_{j-1}$, and thus the total sum is $S(j-1, \ell)$.
\end{itemize}
These two scenarios then result in the following recursion:
\begin{equation}   \label{Eq:sub_problem-recursion}
S(j,\ell) = S(j-1, \ell-1) \cdot f_j + S(j-1, \ell).
\end{equation}
The base cases for the recursion are 
$S(j,j)= \prod_{v \in A_j} f_v$ for $j=1,\ldots,k$ and 
$S(j, 0) = 1$ for $j =1,\ldots,n_i$. If $S(j, 0) = 1$ feels unintuitive, an equivalent substitute set of base cases are $S(j, 1) = \sum_{v \in A_j} f_v$ for $j =1,\ldots,n_i$.

Each step of the recursion takes $O(1)$ time and thus the total time is $O(n_i k + m_i)$ (recall that we assumed all $f_j$ were pre-computed, but that takes $O(m_i)$ time).

\subsection{Estimating $\pi_i$ via Monte Carlo}

In general, network targeting strategies can be quite complicated. We document the following observation in the event that it is useful for considering more complex targeting strategies. Our general approach will be to estimate $\pi_i$ using a modest sample of $R$ probabilities---probabilities with replacement---for seed sets within the sum $\pi_i$ that we can sample uniformly from sets of unique elements (sets {\it without} replacement). More sophisticated estimators of the probability $\pi_i$ may be useful in particularly complex domains. 

Individual probabilities with replacement are often not hard to compute. The difficult in computing $\pi$ is that there are so many sets. Producing a uniform sample can be done easily in $O(km)$ time, and as long as each sample's probability under a complex design is tractable---for one-hop targeting  it takes $O(km)$ time---this computation that follows is a modest $O(kmR)$.

Let $X_1,...,X_R$ be the i.i.d.~random variables representing that probability computed for each sample. Then $\P(X_\ell = p_j) = 1/N$, for each $X_\ell$, each $j$. Consider the estimator 
\begin{eqnarray}
\hat \pi_i = \frac{N}{R} \sum_{i=1}^R X_i.
\end{eqnarray}
If we denote by $p_j$ the constituent probabilities of the sum $\pi_i$, it is clear that 
$$\mathbb E [ \hat \pi_i ] = \frac{N}{R} R \frac{1}{N} \sum_{j=1}^N p_j = \pi_i.$$

Since the estimator $\pi$ relies in $R$ samples from the uniform distribution over sets of size $k$ with unique elements, a natural question is if we can improve the efficiency of the estimator, specifically by using stratified sampling of the seed sets. We briefly give a simple stratification technique (that also has computational advantages). 
Rather than sample $R$ sets each of size $k$, a simple way to sample sets is to take a uniform shuffle of the node order and split the set sequentially into sets of size $k$, producing $r= \lceil m/k \rceil $ sets from a single shuffle. These $r$ sets all constitute unbiased, albeit dependent, samples from the uniform distribution. Their dependence is a useful one, effectively serving as a stratified sample, requiring each node to appear at least once in each set of samples.

We suppress the subscript $i$, studying $\hat \pi$ for a generic village. The variance of $\hat \pi$ is then:
\begin{eqnarray}
\Var [ \hat \pi ] &=& \frac{N^2}{R^2} \sum_{i=1}^R \Var [X_i] \\
&=& \frac{N^2}{R} \frac{1}{N} \sum_{j=1}^N (p_j - \bar p)^2, \label{eqn:pivar}
\end{eqnarray}
where $\bar p = \frac{1}{N} \sum_{j=1}^N p_j$. Notice $N \bar p = \pi$, so if we knew $\bar p$ then we wouldn't need $\hat \pi$. 
As a trivial observation, if the probabilities $p_j$ are uniformly $1/N$ then the variance of $\hat \pi$ is zero. More interestingly, we can bound this variance in non-vacuous ways without estimating it.

{\bf Variance estimate.}
Let $S_p^2$ be the unbiased sample variance of $X_1,\ldots,X_R$, the $R$ probabilities corresponding to uniform samples,
\begin{eqnarray}
S_p^2 = \frac{1}{R-1} \sum_{i=1}^R (X_i - \bar X)^2.
\end{eqnarray}
This sample variance is an unbiased estimate of the variance of each of the i.i.d.~$X_i$'s. As such, we can estimate the variance of $\hat \pi$ as:
\begin{eqnarray}
\widehat{\Var} [ \hat \pi ]
&=& \frac{N^2}{R} S_p^2.
\end{eqnarray}

\newcommand{\pmax}{p_{\text{max}}}
\newcommand{\pmin}{p_{\text{min}}}

{\bf Variance bound. }
The term $\frac{1}{N} \sum_{j=1}^N (p_j - \bar p)^2$ in \eqref{eqn:pivar} is the variance of a discrete random variable that has compact support on $[0,1]$, and we can tighten this support further by deriving maximum and minimum probabilities $\pmax$ and $\pmin$. For a discrete random variable with compact support on $[\pmin, \pmax]$, we can bound \eqref{eqn:pivar} by:
\begin{eqnarray}
\Var [ \hat \pi ] &=& \frac{N^2}{R} \frac{1}{N} \sum_{j=1}^N (p_j - \bar p)^2 \\
& \le & \frac{N^2}{R} \frac{1}{2} \left [ \left (\pmax - \frac{\pmax + \pmin}{2} \right )^2 + \left (\pmin - \frac{\pmax + \pmin}{2} \right )^2 \right ] \\
& = & \frac{N^2}{4R} (\pmax - \pmin)^2. \label{eqn:pivarbound1}
\end{eqnarray}
In this bound the term $N^2$ is typically enormous, but $\pmax^2$ is often proportionally small. As a result, whenever one can upper bound the maximum probability of a set being selected, such a bound can be translated to a bound on the variance of $\hat \pi$ at a given $R$.

\newpage
\section{Effective sample size diagnostics}
\label{app:ess}
A measure of effective sample size, obtained by comparing the variance to standard unweighted averages, can be a useful diagnostic for determining whether the importance distribution carries enough information to estimate the mean of the target distribution and may inform whether or not asymptotic approximations for inference are appropriate~\citep[Section 9.3]{mcbook}. We focus on the self-normalized (H\'ajek) estimator in this section, as it is the weighted equivalent of an unweighted sample average and is our preferred estimator.  

\subsection{Off-policy population mean}
\label{app:ess-op-mean}
Consider the off-policy setting where we have observed outcomes under an importance distribution $p_B$ and wish to estimate the population mean under a target distribution $p_A$.  This is the case, for example, when seed sets are assigned according to random targeting ($B$) but we wish to make inferences about one-hop targeting ($A$).  Given weights $w_i^A = a_i / b_i$ as in equation~\eqref{eqn:weights}, that mean $\mu_A$ is estimated using the weighted average
\[\hat \mu_A = \frac{\sumin w_i^A y_i}{\sumin w_i^A} \]
Now, let $\sigma^2 = \var_A(y_i(S_i))$. This imposes homoskedasticity; one natural way for this to be satisfied is under the sharp null that outcomes are not affected by the seed set, equation \eqref{eqn:sharp_null}.
Then, conditionally on the observed weights,
\[\var(\hat \mu_A) = \sigma^2 \frac{\sumin (w_i^A)^2}{(\sumin w_i^A)^2} = \sigma^2 \frac{\overline{w_A^2}}{n\bar w_A^2},\]
where
\[\bar w_A = \avgin w_i^A, \qquad \overline{w_A^2} = \avgin (w_i^A)^2.\]
In contrast, an unweighted sample average of $\neff$ independent observations has variance $\sigma^2 / \neff$, so $\hat \mu_A$ has the same variance as an unweighted average of
\begin{equation}
\label{eqn:neff-one-sample}
\neff = \frac{(\sumin w_i^A)^2}{\sumin (w_i^A)^2} = \frac{n \bar w_A^2}{\overline{w_A^2}}
\end{equation}
observations.  If $\neff$ is much smaller than $N$, then it may be the case that $p_A$ is too different from $p_B$ to be able to estimate $\mu_A$ using observations from $p_B$.  This notion of defining an effective sample size for weighted averages is quite old and is also known as a \emph{design effect} in the survey sampling literature~\citep[see, for example,][]{kish1965survey}.

The $w_i^A$ are i.i.d.~observations of a random variable $W_A = w_A(S)$, having population moments 
$\E_A[\bar w_A] = 1$ and $\E_A[\overline{w_A^2}] = \E_A W_A^2 = \E_B W_A$.  So a population version of $\neff$ can be written as 
\begin{equation}
\neff^* = \frac{N (\E_A[\bar w_A])^2}{\E_A[\overline{w_A^2}]} = \frac{N}{\E_A W_A^2} = \frac{N}{\E_B W_A}.
\label{eqn:neff-off-policy-population}
\end{equation}
If the seed set distributions are known in advance then $\neff^*$ can be computed prior to launching a field experiment, which can give some indication of the informativeness of the experiment, say, with respect to different counterfactual seeding strategies.  This may be useful when the entire social network is observed and when the seed sets are of small enough size to permit computation of the expectation specified in the expression for $\neff^*$.  Otherwise a Monte Carlo estimate of $\neff^*$ can easily be constructed by sampling seed sets, or the sample version $\neff$ can be used instead.

\subsection{Average treatment effect}
\label{app:ess-ate}
Now consider an experiment designed to compare strategies $A$ and $B$, with observations assigned to both strategies, as in the field experiment conducted by \citet{kim2015social}.
Consider the H\'ajek estimator $\tilde \tau$ of the average treatment effect
\[
\tilde \tau = \avgin\left(\frac{w_i^A}{\bar w_A} - \frac{w_i^B}{\bar w_B}\right)y_i.
\]
As in the off-policy case, if we consider the weights as fixed, then the variance is the sum of squares of the weights,
\[
V_{\tilde \tau} = 
\frac{\sigma^2}{N^2} \sumin \left(\frac{w_i^A}{\bar w_A} - \frac{w_i^B}{\bar w_B}\right)^2  = \frac{\frac{1}{n}\sumin (\bar w_B w_i^A - \bar w_A w_i^B)^2}{n\bar w_A^2 \bar w_B^2}. 
\]

In contrast to the off-policy case, there is no standard notion of effective sample size for this two-sample scenario.  The appropriate point of comparison is less clear because we are estimating a difference between two means.  One possibility is to use as a comparison an ordinary two-sample equal-variance difference-in-means estimator.

Now to construct a baseline comparison estimator, consider a hypothetical setup where we run an ordinary experiment.  We assign $n_A$ observations independently to $A$ and $n_B = N - n_A$ observations to $B$, and use weights
\begin{align*}
w_i^A &= \begin{cases}
1 &\text{if }i = 1, \dots, n_A \\
0 &\text{if }i = n_A + 1, \dots, N
\end{cases} \\
w_i^B &= 1 - w_i^A.
\end{align*}
Then the H\'ajek estimator reduces to the difference-in-means estimator and the variance is
\[
V_{\DM} = 
\frac{\sigma^2}{N^2} \sumin \left(\frac{w_i^A}{\bar w_A} - \frac{w_i^B}{\bar w_B}\right)^2 = \frac{\sigma^2}{N^2} \left(\frac{n_A}{\bar w_A^2} + \frac{n_B}{\bar w_B^2} \right) = \sigma^2 \left(\frac{1}{n_A} + \frac{1}{n_B}\right),
\]
the variance of an ordinary two-sample difference-in-means.  For example, if our experiment consists of data from 100 villages where 50 villlages are assigned to random targeting and 50 villages are assigned to one-hop targeting, then the difference-in-means (DM) estimator has variance $\sigma^2 \times (1 / 50 + 1 / 50) = \sigma^2 / 25$.

One reasonable definition of effective sample size, then, is to scale the original sample size $N$ by the ratio of these two variances,
\begin{equation}
\label{eqn:neff-two-sample}
\neff = 
\underbrace{N}_{\text{original sample size}} \times
\underbrace{\left(\frac{1}{n_A} + \frac{1}{n_B}\right)}_{\text{DM scaling}} \times 
\underbrace{\frac{N\bar w_A^2 \bar w_B^2}{\frac{1}{N}\sumin (\bar w_B w_i^A - \bar w_A w_i^B)^2}}_{\text{inverse H\'ajek variance}}.
\end{equation}
Of course, this is the same as 
\begin{equation}
\label{eqn:ess-ate}
\neff = 
\frac{1}{\rho(1 - \rho)} \times 
\frac{N\bar w_A^2 \bar w_B^2}{\frac{1}{N}\sumin (\bar w_B w_i^A - \bar w_A w_i^B)^2},
\end{equation}
where $\rho$ is the proportion of units assigned to group $A$. In essence, $\neff$ is defined exactly so that the difference-in-means estimator in a Bernoulli$(\rho)$ experiment has $\neff = N$ (and should thus be viewed conditionally on the proportion parameter $\rho$).  Note that $n_A$, $n_B$, and $\rho$ are parameters for the comparison Bernoulli experiment, and have nothing to do with the distribution generating the observed data.

As $\E[\bar w_A] = \E[\bar w_B] = 1$, the denominator on the right-hand side of equation~\eqref{eqn:neff-two-sample} is a plug-in estimator for $\E[(W_A - W_B)^2]$.  
Therefore, a population version of $\neff$ can be stated as
\begin{equation}
\neff^* = \frac{1}{\rho(1 - \rho)}\times \frac{N}{\E[(W_A - W_B)^2]} = \frac{1}{\rho(1 - \rho)} \times \frac{N}{\E\left[\left(\frac{P_A - P_B}{P_\Delta}\right)^2\right]}.
\label{eqn:neff-two-sample-population}
\end{equation}
Notice that the effective sample size depends on the targeting distributions only through $\E\left[\left(\frac{P_A - P_B}{P_\Delta}\right)^2\right]$, capturing the intuition that the $\neff^*$ is lowered if discordant seed sets (those with very different probabilities between the two targeting strategies) are not accounted for by the design.  Evaluating $\neff^*$ can be a useful indicator for how powerful hypothesis tests for comparing strategies $A$ and $B$ may be, and can be done before running any experiments provided the network structures are known. 

\subsection{Off-policy average treatment effect}
\label{app:ess-op-ate}
As a special case of Appendix~\ref{app:ess-ate}, we consider the case where all villages are assigned to a single seeding strategy (say strategy B, random targeting).  This is distinct from the scenario considered in Appendix~\ref{app:ess-op-mean} because we consider the estimand to be the treatment effect rather than simply the off-policy population mean.  In this case $w_i^B = \bar w_B = 1$, so equation~\eqref{eqn:ess-ate} becomes
\begin{equation}
\label{eqn:neff-op-ate}
\neff = \frac{1}{\rho(1 - \rho)}\times \frac{N \bar w_A^2}{\frac{1}{N} \sum_{i=1}^N (w_i^A - \bar w_A)^2}.
\end{equation}
Since $\frac{1}{N}\sum_{i=1}^N (w_i^A - \bar w_A)^2$ is the sample version of $\var_A(W_A) = \E_A(W_A)^2 - \E_A(W_A) = \E_A(W_A)^2 - 1 = \E_B W_A - 1$,
a population effective sample size is
\begin{equation}
\label{eqn:neff-op-ate-population}
\neff^* = \frac{1}{\rho(1 - \rho)}\times \frac{N}{\E_B W_A - 1}.
\end{equation}
Note the difference between this $\neff^*$ and the derivation in equations~\eqref{eqn:neff-one-sample} and~\eqref{eqn:neff-off-policy-population} given in Appendix~\ref{app:ess-op-mean}; there, the estimand is the mean under strategy $A$ and equation~\eqref{eqn:neff-one-sample} corresponds to the standard measure of effective sample size (or {\it design effect}) used in importance sampling.  By contrast, equations~\eqref{eqn:neff-op-ate} and~\eqref{eqn:neff-op-ate-population} are for estimating the treatment effect between the two strategies. The estimators have different variances and hence different effective sample sizes.

\newpage
\section{Optimizing the experimental design}
\label{app:opt-design}
For experiments specifically seeking to evaluate the difference between stochastic seeding strategies, the variance expressions for the Horvitz--Thompson and H\'ajek estimators, equations \eqref{eqn:ht-variance} and \eqref{eqn:hajek-variance} respectively, can be considered as objective functions of an optimization problem that tries to optimize the power of the experiment. Our approach to this problem is based on knowing the structure of the networks where the evaluation is taking place, in line with the network data collection process for the \citet{kim2015social} study that preceded the rollout of the experiment. The procedure below is not necessarily an optimal experimental design -- a genuinely optimal design would need information about the responses $Y$ -- which is why we refer to this design strategy as an {\it optimized} design rather than an optimal design.
 
  Note that all of the analysis in this section relies on the fact that $Y$ is non-negative; otherwise, one may treat the positive and negative parts of $Y$ separately.
  
First consider the Horvitz--Thompson estimator.  Following from equation~\eqref{eqn:ht-variance}, a good choice of $P_\Delta$ is one that minimizes
\[\sigma_\Delta^2 := \E\left[\frac{(P_A - P_B)^2Y^2}{P_\Delta^2}\right] - \tausp^2 = \E\left[\frac{(p_A(S) - p_B(S))^2y(S)^2}{p_\Delta(S)^2}\right] - \tausp^2.\]
Then the choice
\[p_{\Delta^*} \propto \frac{|(p_A(S) - p_B(S))y(S)|}{\tausp}\]
is optimal.  To see this, let $p_\Delta$ be any other design.  Then 
\begin{align*}
\sigma_{\Delta^*}^2 + \tausp^2 &= \E_{\Delta^*} \left[\frac{(p_A(S) - p_B(S))^2y(S)^2}{p_{\Delta^*}(S)^2}\right] = \tausp^2 \\
&= \E_\Delta \left[\frac{(p_A(S) - p_B(S))y(S)}{p_\Delta(S)}\right]^2 \leq \E_\Delta \left[\frac{(p_A(S) - p_B(S))^2y(S)^2}{p_\Delta(S)^2}\right] = \sigma_\Delta^2 + \tausp^2,
\end{align*}
using the Cauchy-Schwarz inequality.  Of course, $p_{\Delta^*}$ is not actually a useful distribution since it depends on the unknown  true treatment effect, but it can provide hints on how to proceed.  In particular, it suggests that seed sets for which the difference in probabilities $|P_A - P_B|$ and the magnitude of the response $|Y|$ are both large provide the most information about the hypothesis, and the distribution of the experimental design should thus place more weight on these seed sets.

Since the target distributions $P_A$ and $P_B$ are known, we must proceed by making assumptions about the response $Y$.  The simplest such assumption is to assume that $y(S)$ is constant across sets $S$, giving us the design distribution 
\begin{equation}
\label{eq:app-pdelta}
p_{\Delta^*} \propto |P_A - P_B|.
\end{equation}
This approach would maximize power for the Horvitz--Thompson estimator in the event that the true data generating process is invariant to the choice of seed set and, thus, to choice of seeding strategy.  

Under this optimized design, the design probability of a seed set will be zero whenever the probability of the set under one-hop targeting equals the probability under uniform random targeting, so it satisfes Assumption~\ref{assumption:positivity}.  However, it need not satisfy corresponding positivity assumptions for $P_A$ and $P_B$ separately. Thus, the difference will be identified even if the means $\mu_A$ and $\mu_B$ are not.
In practice, researchers might want to ensure that all seed sets have positive probability in case other seeding strategies become interesting ex post or it is desirable to estimate the means as well.

A more sophisticated approach would model the response using domain knowledge, perhaps via a social influence model such as the independent cascade model or the linear threshold model~\citep{kempe2003maximizing}.  The most reliable approach might be to use the results of a previous, pilot experiment to inform the design of the next experiment; such a procedure is a form of adaptive importance sampling~\citep[Section 10.5]{mcbook}. 

Designing an optimized experiment for the H\'ajek estimator is similar to the Horvitz--Thompson procedure.  Examining the variance expression in equation~\eqref{eqn:hajek-variance}, an optimized design is given by
\[
p_{\Delta^*} \propto |\mu_A p_A(S) - \mu_B p_B(S) - y(S)(p_A(S) - p_B(S))|.
\]
Again, $p_{\Delta^*}$ relies on the unknown quantities $Y$, $\mu_A$ and $\mu_B$, which must either be subjected to strong assumptions (e.g., assumed to be constant) or estimated in some way using historical data or domain knowledge. For our evaluations in Section~\ref{sec:design_and_ess}, where we analyze the population effective sample size of experiments under Bernoulli(0.5) designs compared to designs optimized for the H\'ajek estimator, we use the design distribution in equation~\ref{eq:app-pdelta}. This design distribution, which was minimizing the variance of the Horvitz--Thompson estimator when $Y$ is constant, can also be thought of as minimizing the variance of the H\'ajek estimator when $\mu_A = \mu_B \ne 0$, but $y(S)$ is not constant across sets $S$ or is constant across seed sets $S$ but varies between units.

\clearpage

\section{Supplementary simulations}
\label{app:simulations}

Here we report on additional simulations. This includes some additional generative models, and also some variations on the sample size. Furthermore, some additional analyses of the simulations in the main text are reported here.

\subsection{Additional analyses and larger treatment effects}
\label{app:larger_effects}
In order to allow for larger differences between one-hop and random seeding, we modify the probit model in Equation \ref{eqn:sim-model} to allow for additional, direct effects of the degree of the seed set on adoption. This might reflect that individuals are trying to coordinate with higher degree individuals, whether or not they are themselves connected to them; however, we impose this simply as a parsimonious way of amplifying treatment effects to stress-test estimation and inference. In particular, we define the $t$-th time step response using the probit model
\begin{align}
Y_{ij,t}^* &= \alpha + \beta Z_{ij,t} + \gamma X_i + \ep_{ij,t}, \label{eqn:sim-model-degree}\\
Y_{ij,t} &= \max\{Y_{ij,t-1},\one(Y_{ij,t}^* > 0)\}.\nonumber
\end{align}
This adds the term $\gamma X_i$ where $X_i$ is the summed in-degrees of the seed set for village $i$; that is, $X_i = \sum_{j \in S_i} d^{+}_{ij}$. We report simulations here at include setting $\gamma=0.1$, in addition to those in the main text, which set $\gamma=0$.
Including this feature allows us to further vary the treatment effect between one-hop and random targeting in our simulations because one-hop targeting generates high degree seeds more often than random targeting.

\begin{figure}[bt]
\centering
\includegraphics[width=\textwidth]{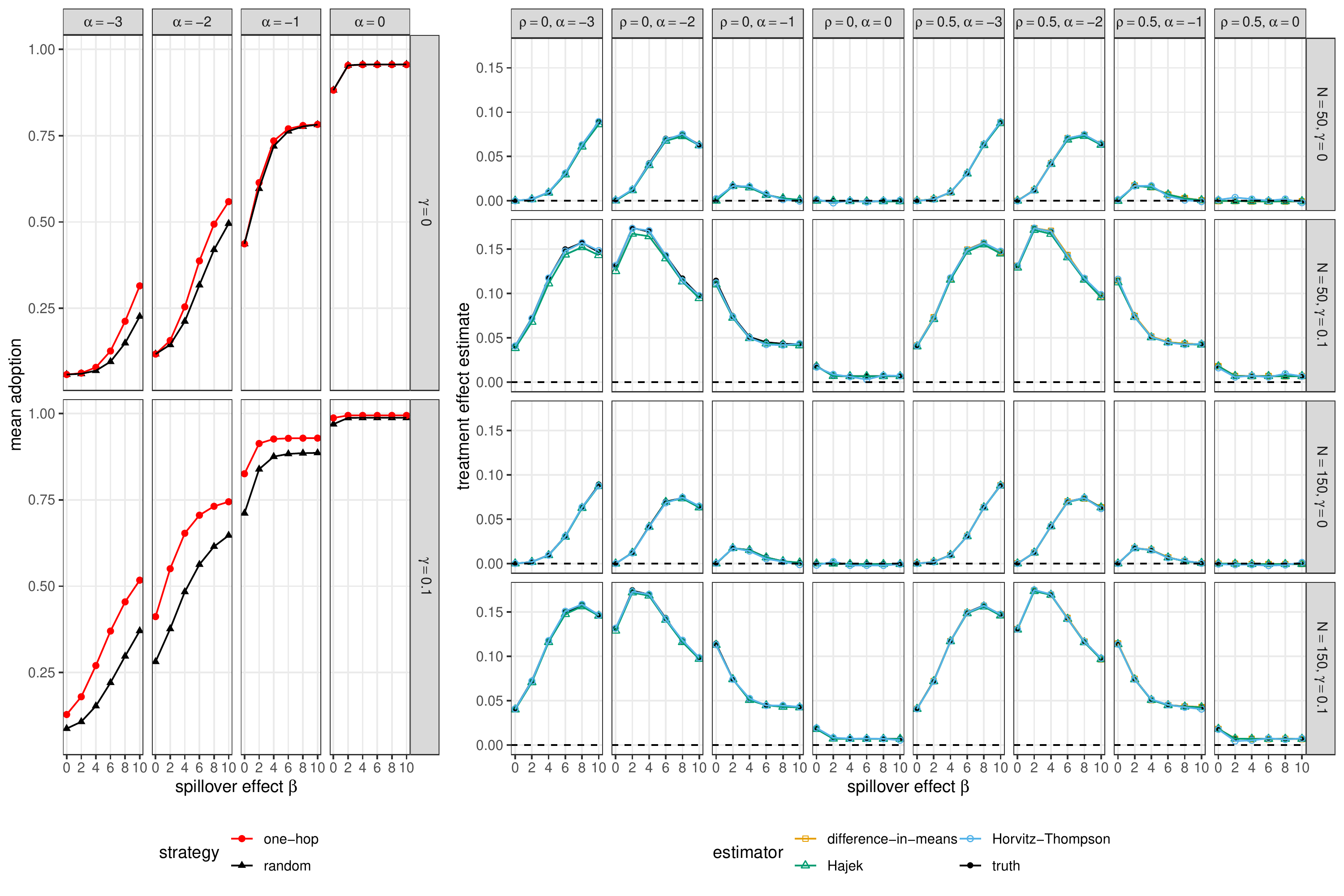}
\caption{(left) True mean adoption rates for random and one-hop targeting for the simulation setup described in Section~\ref{app:larger_effects}. (right) True treatment effect and estimated values from the difference-in-means, Horvitz--Thompson, and H\'ajek estimators.  Columns vary the intercept $\alpha$ and probability of assignment to one-hop $\rho$. Rows vary the number of villages $N$ and the coefficient for summed in-degrees of the seed set $\gamma$. The horizontal axis varies the spillover effect $\beta$. As intended, increasing $\gamma$ results in larger effects to more fully examine performance of estimation and statistical inference. Figure \ref{fig:sim-true-tau} in the main text thus displays a subset of the panels shown here.}
\label{fig:sim-true-tau-all}
\end{figure}

\begin{figure}[bt]
\centering
\includegraphics[width=\textwidth]{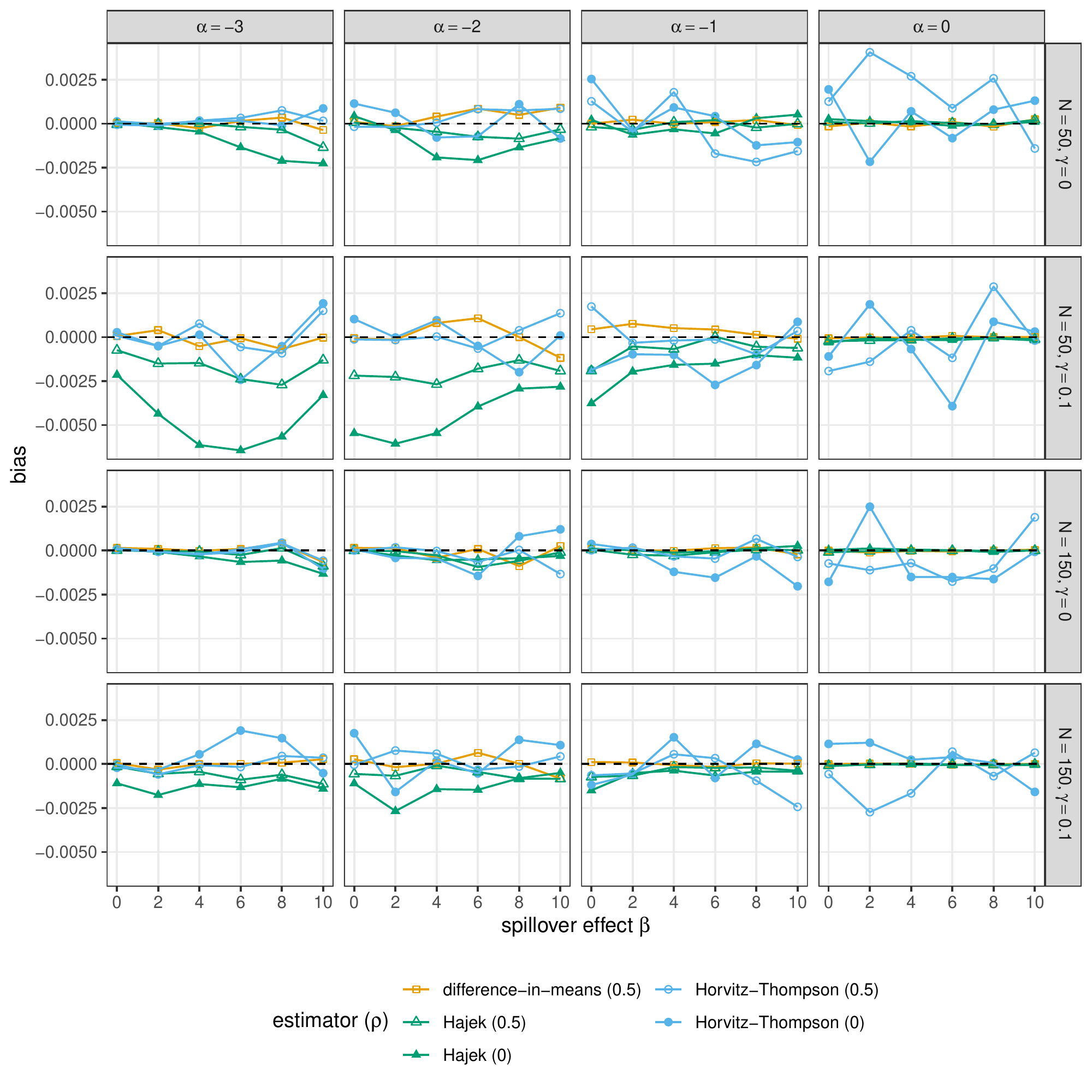}
\caption{Estimated bias of estimators in the simulation setup described in Section~\ref{app:larger_effects}.
Columns vary the intercept $\alpha$. Rows vary the number of villages $N$ and the coefficient for summed in-degrees of the seed set $\gamma$. The horizontal axis varies the spillover effect $\beta$. Probability of assignment to one-hop is varied between $\rho = 1/2$ (unfilled, as also shown in the main text) and $\rho = 0$, i.e., off-policy (filled).
Note that the difference-in-means estimator is not defined off-policy, so it not shown for the design without any one-hop villages (i.e., $\rho = 0$).
In smaller sample sizes with rare outcomes, the H\'ajek estimator exhibits some downward bias in off-policy estimation.
}
\label{fig:sim-bias-all}
\end{figure}

\begin{figure}[bt]
\centering
\includegraphics[width=\textwidth]{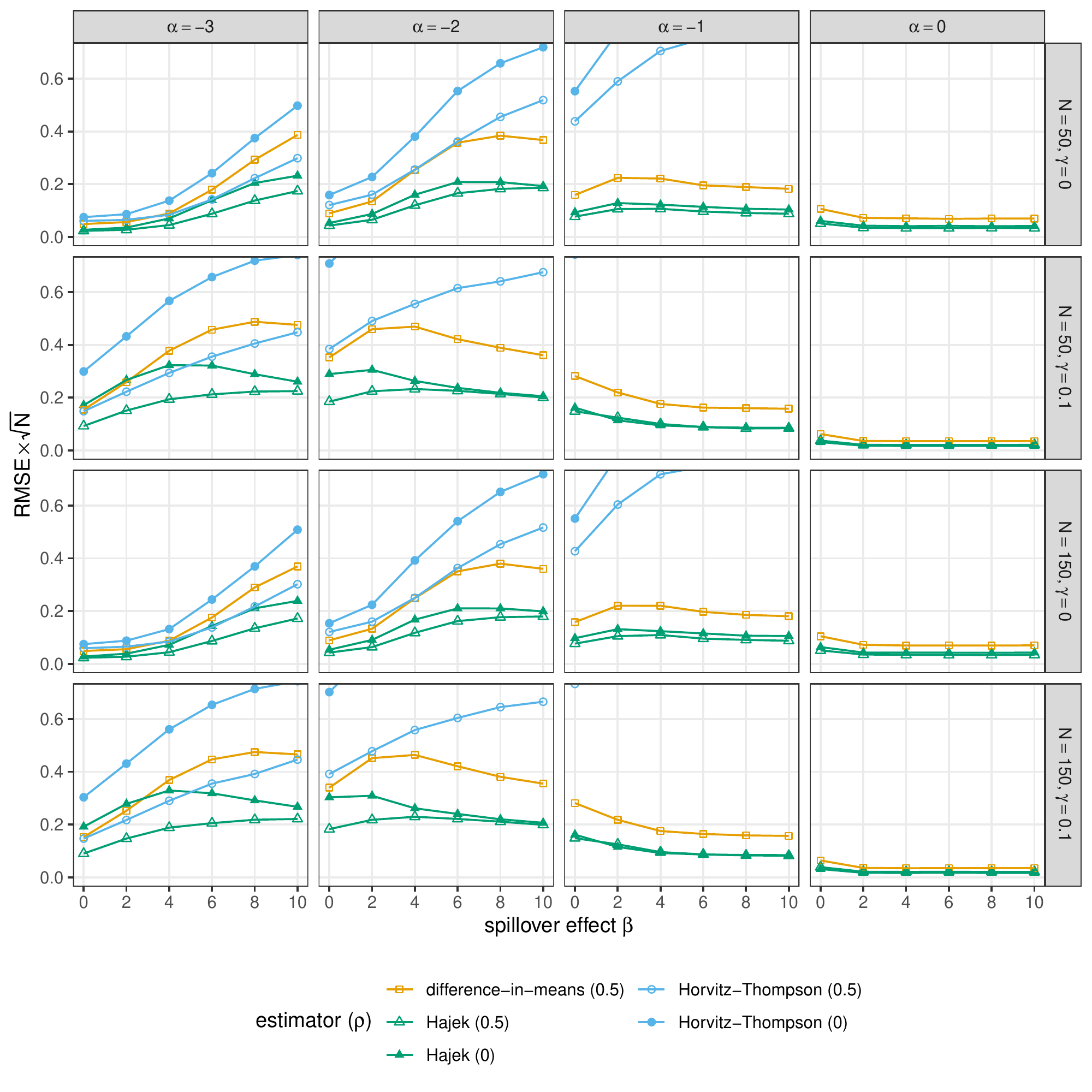}
\caption{Estimation error (RMSE $\times \sqrt{N}$) in the simulation setup described in Section~\ref{app:larger_effects}.
Columns vary the intercept $\alpha$ and probability of assignment to one-hop $\rho$. Rows vary the number of villages $N$ and the coefficient for summed in-degrees of the seed set $\gamma$. The horizontal axis varies the spillover effect $\beta$.
Note that the difference-in-means estimator is not defined ``off-policy'', so it not shown for the design without any one-hop villages (i.e., $\rho = 0$).
In some cases, the Horvitz--Thompson has very high variance, as expected, and so its RMSE curve is not visible so as to keep the scale informative for other comparisons.
}
\label{fig:sim-rmse-all}
\end{figure}

Figure~\ref{fig:sim-true-tau-all} shows the true outcomes and treatment effects along with mean estimates of treatment effects for the larger space of generative models. As expected, setting $\gamma = 0.1$ produces larger treatment effects.
Figure~\ref{fig:sim-bias-all} shows the estimated bias for all estimators, including in the off-policy design with $\rho = 0$. Except for smaller sample sizes off-policy, all are approximately unbiased.
Figure~\ref{fig:sim-rmse-all} shows the estimation error for all estimators. These results are all qualitatively consistent with those in the main text, with the proposed H\'ajek performing best.

\begin{figure}[h]
\centering
\includegraphics[width=\textwidth]{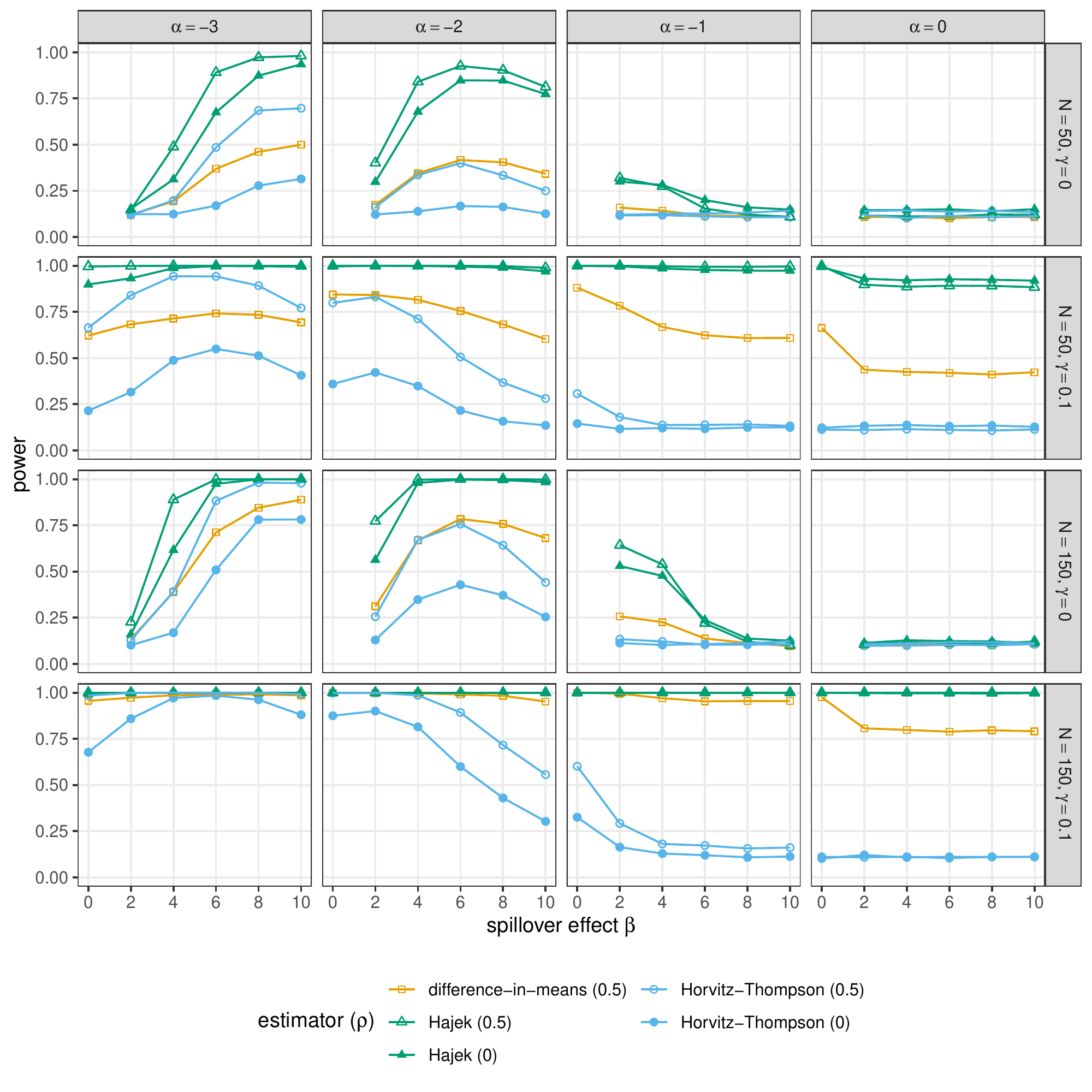}
\caption{Power (rejection rates when the null is false) of estimators and associated analytical hypothesis tests in the simulation setup described in Section~\ref{app:larger_effects}.
Note that some of these curves correspond to tests that are anticonservative (i.e., their corresponding confidence intervals have below-nominal coverage), particularly when $N = 50$ and in the off-policy design $\rho = 0$.
}
\label{fig:sim-power}
\end{figure}

Figure~\ref{fig:sim-power} plots the power of the experiment (the fraction of experiments in which the null was rejected) as a function of both the simulation parameters (left) and the true treatment effect (top right).  For the response model and parameter distribution used in our simulations, the H\'ajek estimator generally has substantially more power than the difference-in-means estimator, while the Horvitz--Thompson estimator is underpowered due to excessive variance~(Figure~\ref{fig:sim-power}, bottom right).

Figure~\ref{fig:sim-power} plots the power of the experiment (the fraction of experiments in which the null was rejected) as a function of both the simulation parameters (left) and the true treatment effect (top right).  For the response model and parameter distribution used in our simulations, the H\'ajek estimator generally has substantially more power than the difference-in-means estimator, while the Horvitz--Thompson estimator is underpowered due to excessive variance~(Figure~\ref{fig:sim-power}).

\subsection{Independent cascade model}
\label{app:IC}
We also report simulation results using an entirely different model of adoption, the independent cascade model, in which each new adopter has an independent chance $p$ of causing adoption for each of its neighbors \citep{kempe2003maximizing}. The independent cascade model is widely used in work on seeding, especially in computer science, and it is a special case of a susceptible--infectious--recovered (SIR) compartmental model in discrete time when transition from the infectious to recovered state is deterministic. In particular
$$
\Pr(Y_{ij,t} | Y_{ij,t-1} = 0) = 1 - (1 - p)^{V_{ij, t}}
$$
where $V_{ij,t} = \sum_k G_{ijk} |Y_{ik,t-1} - Y_{ik,t-2}|$ is the number of neighbors who first adopted in the previous time-step. We run this process until no new adoptions occur. For each combination of simulation parameters, we conduct 1,000 replications.

\begin{figure}[bt]
\centering
\includegraphics[width=.7\textwidth]{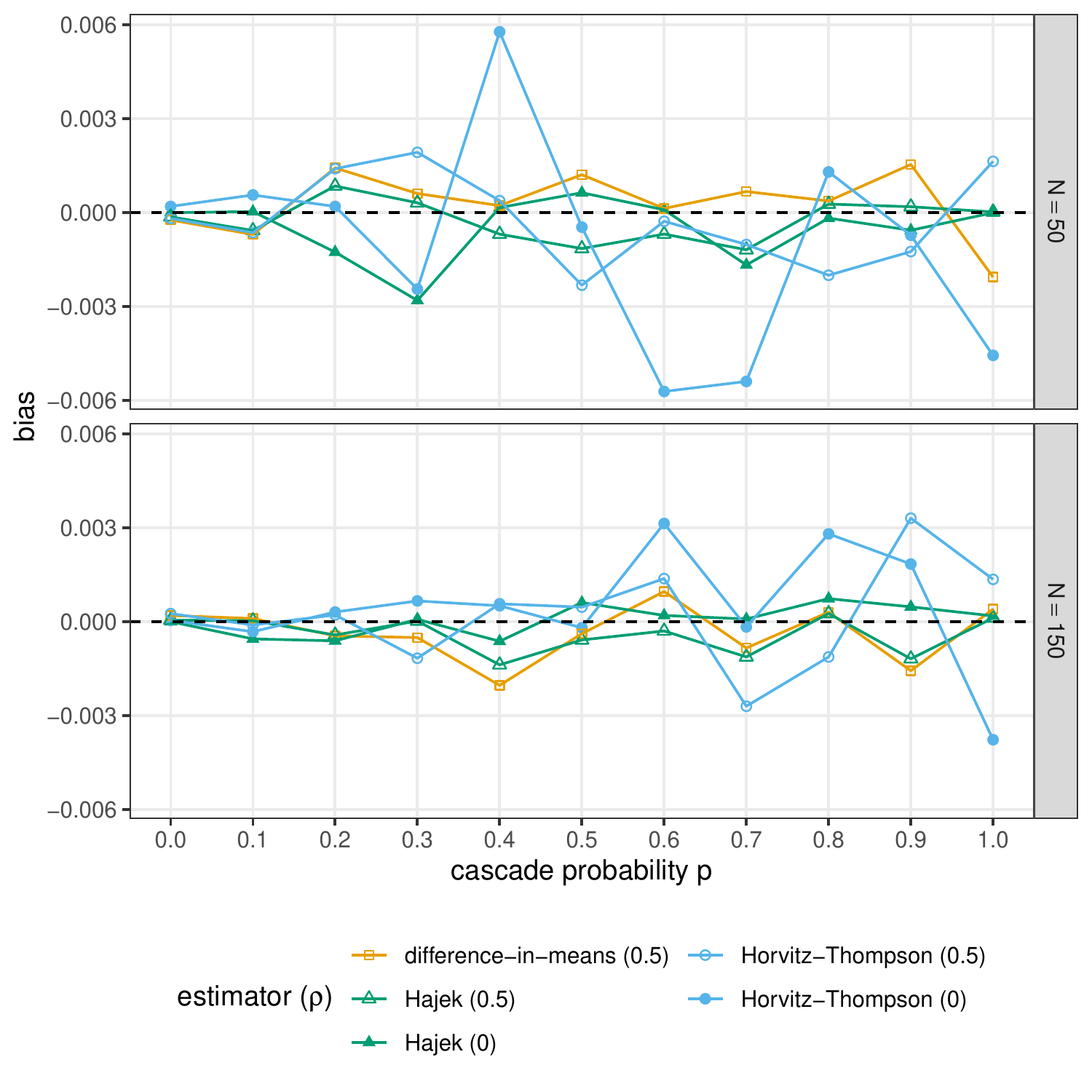}
\caption{Estimated bias of estimators in the independent cascade simulation setup described in Section~\ref{app:IC}.
Rows vary the number of villages $N$. The horizontal axis varies the cascade probability $p$. Probability of assignment to one-hop is varied between $\rho = 1/2$ (unfilled) and $\rho = 0$, i.e., off-policy (filled).
Note that the difference-in-means estimator is not defined off-policy, so it not shown for the design without any one-hop villages (i.e., $\rho = 0$).
}
\label{fig:sim-IC-bias}
\end{figure}

\begin{figure}[bt]
\centering
\includegraphics[width=.7\textwidth]{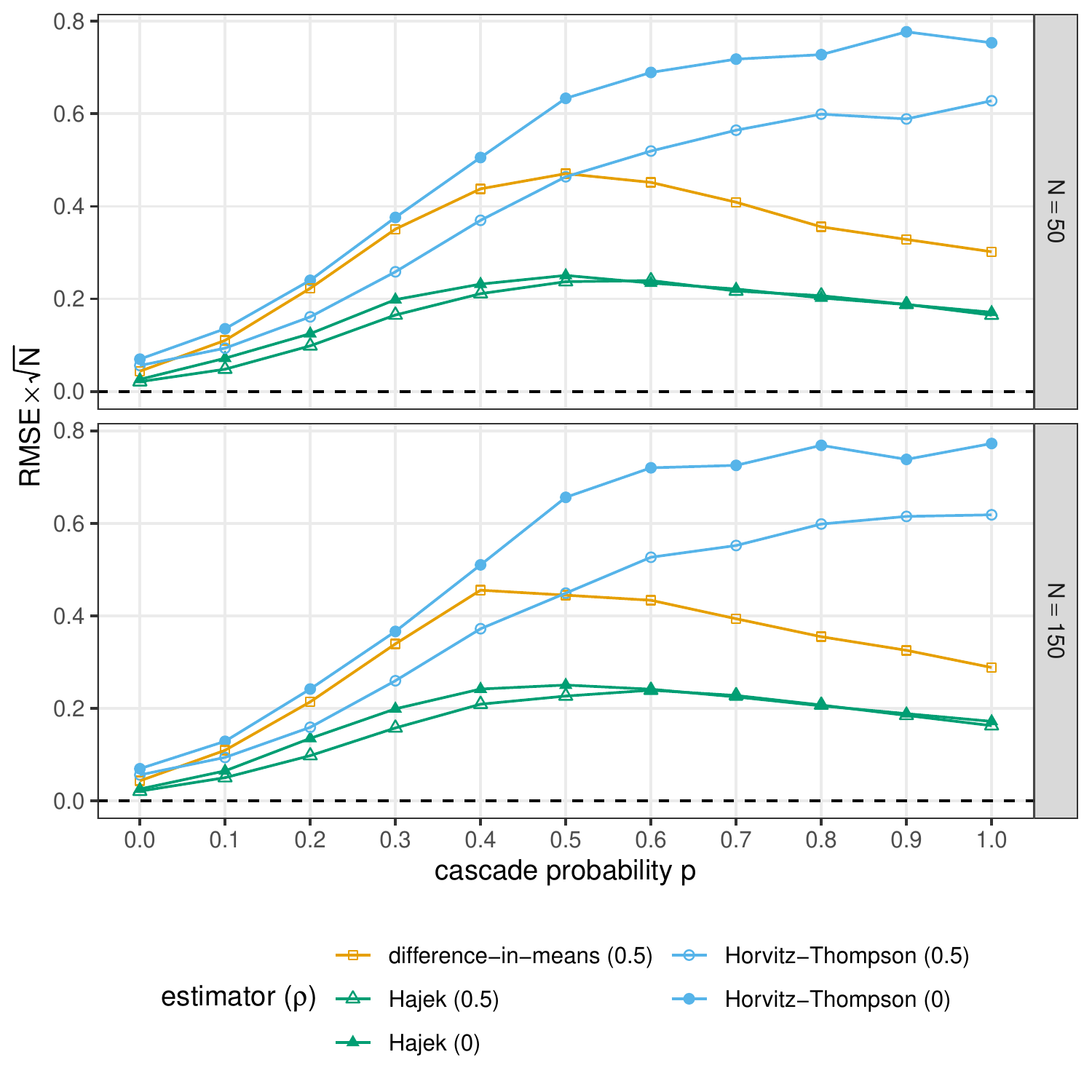}
\caption{Estimation error (RMSE $\times \sqrt{N}$) in the independent cascade simulation setup described in Section~\ref{app:IC}.
Rows vary the number of villages $N$. The horizontal axis varies the cascade probability $p$. Probability of assignment to one-hop is varied between $\rho = 1/2$ (unfilled) and $\rho = 0$, i.e., off-policy (filled).
Note that the difference-in-means estimator is not defined off-policy, so it not shown for the design without any one-hop villages (i.e., $\rho = 0$).
}
\label{fig:sim-IC-rmse}
\end{figure}

\begin{figure}[bt]
\centering
\includegraphics[width=.7\textwidth]{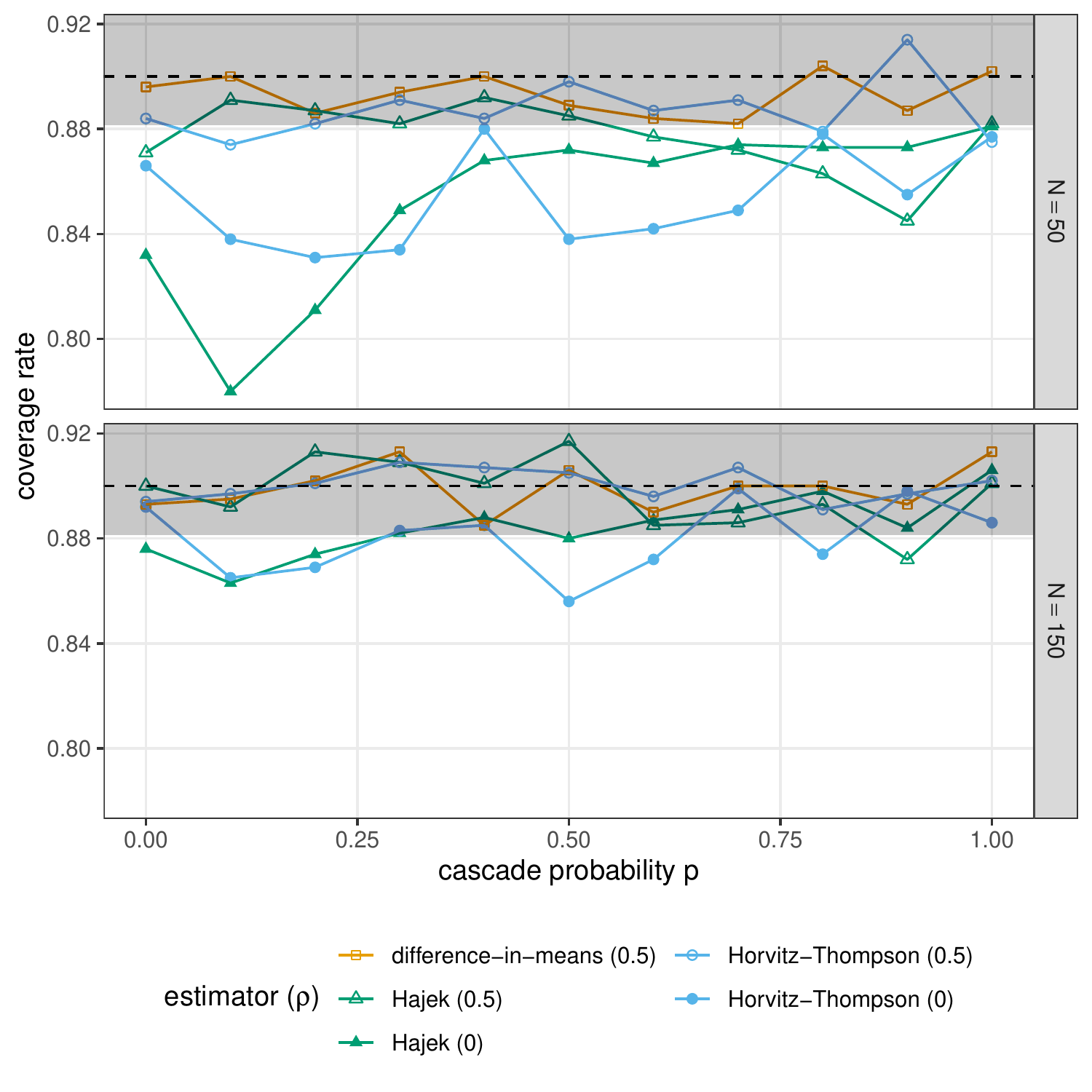}
\caption{Coverage rates for a 90\% nominal confidence interval under the independent cascade simulation setup described in Section~\ref{app:IC}; shaded area is the 95\% acceptance region ($p > 0.05$) for coverage being at least the nominal rate.
Rows vary the number of villages $N$. The horizontal axis varies the cascade probability $p$. Probability of assignment to one-hop is varied between $\rho = 1/2$ (unfilled) and $\rho = 0$, i.e., off-policy (filled).
Note that the difference-in-means estimator is not defined off-policy, so it not shown for the design without any one-hop villages (i.e., $\rho = 0$).
All estimators have approximately nominal coverage with $N=150$, with $N=50$ inference using this asymptotic approximation is anti-conservative, notably for small sample sizes with a very rare outcome ($\alpha = -3, N = 50$)
}
\label{fig:sim-IC-coverage}
\end{figure}

We present very similar analyses to those in the main text and in Appendix \ref{app:larger_effects}. First, Figure \ref{fig:sim-IC-true-tau} presents the true treatment effects and mean estimates as a function of the cascade probability $p$. Again, all the estimators are approximately unbiased, a point further illustrated in Figure \ref{fig:sim-IC-bias} for both Bernoulli(1/2) and off-policy settings. Figure \ref{fig:sim-IC-rmse} displays the overall estimation error. As with prior results, our H\'ajek estimator results in lower RMSE than the difference-in-means estimator, including when relying on random seeding data only. Figure \ref{fig:sim-IC-coverage} presents the coverage of resulting analytic confidence intervals; as with our other results, there is some undercoverage for $N=50$, here particularly for off-policy evaluation.

\begin{figure}[bt]
\centering
\includegraphics[width=.9\textwidth]{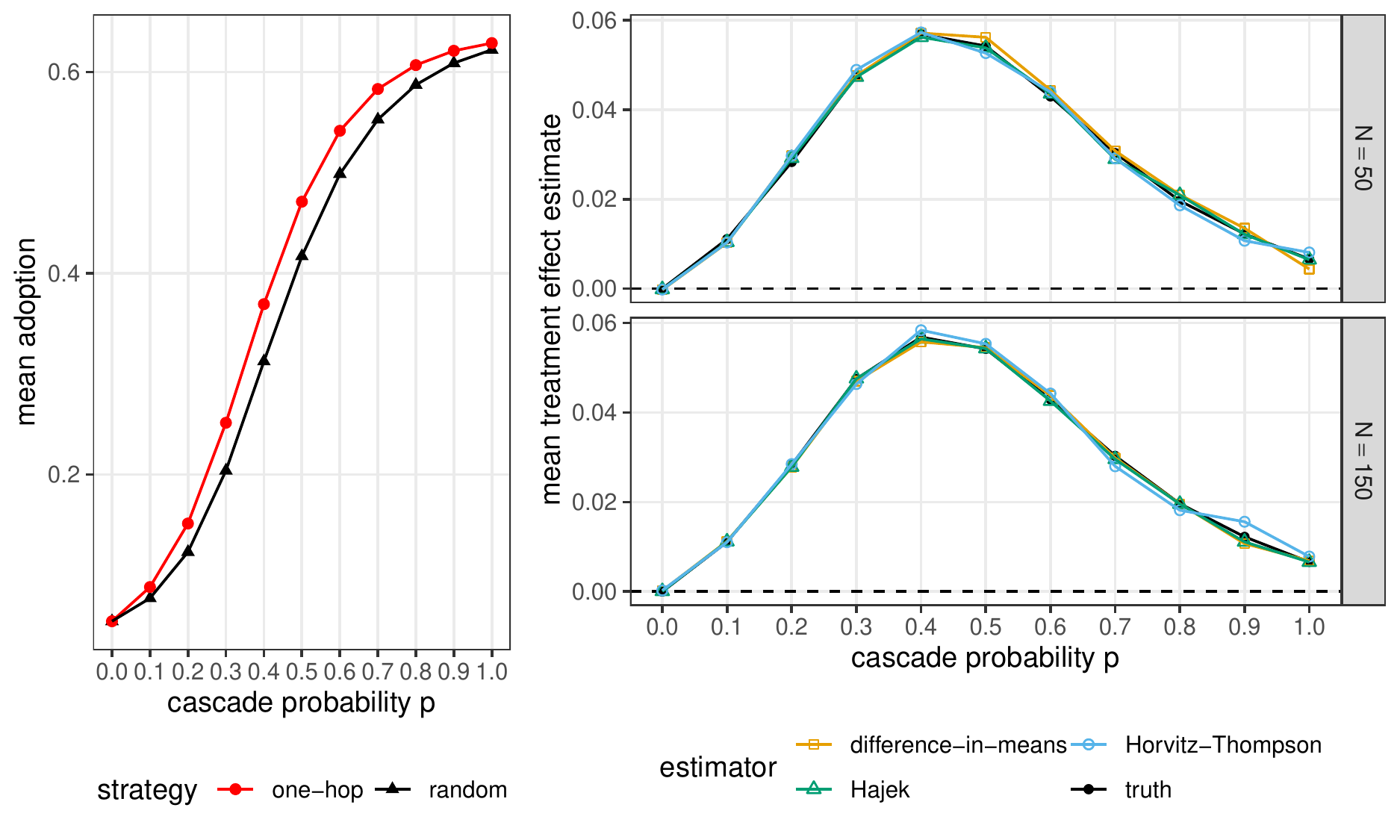}
\caption{(left) True mean adoption rates for random and one-hop targeting for independent cascade simulation setup described in Section~\ref{app:IC}. (right) True treatment effect and estimated values from the difference-in-means, Horvitz--Thompson, and H\'ajek estimators in the Bernoulli($1/2$) design.  Rows vary the number of villages $N$. The horizontal axis varies the cascade probability $p$.}
\label{fig:sim-IC-true-tau}
\end{figure}

\clearpage
\paluck{
\section{Supplementary empirical analyses}
\label{app:empirical}

\subsection{Kim et al.~(2015)}
\label{app:kim}

As discussed in the main text, seed sets selected under random seeding can often have higher mean in-degree than seed sets selected under one-hop seeding. Because of the use of seeding for two different products in \cite{kim2015social}, it is possible to observe how often this happens in the eight villages assigned to one-hop and random seeding for the two products. Figure \ref{fig:kim-seeds} shows the mean in-degree for these eight villages, with three exhibiting this reversal.

\begin{figure}[tb]
\centering
\includegraphics[height=7cm]{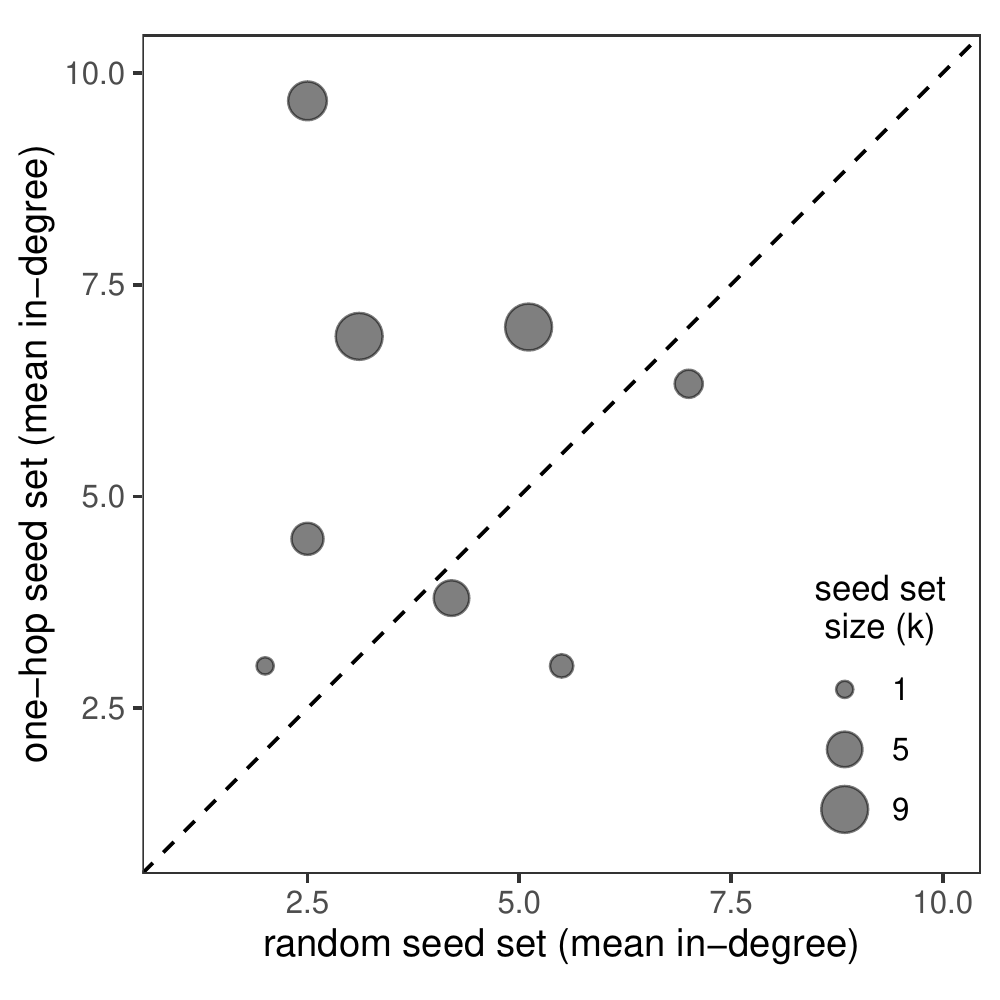}
\caption{Mean in-degree of the seed sets selected in \citet{kim2015social} for those villages assigned to one-hop targeting for one product (multivitamins or chlorine) and random targeting for the other. This is based on data presented in Table S3 of \cite{kim2015social}}
\label{fig:kim-seeds}
\end{figure}

\subsection{Cai et al.~(2015)}
\label{app:cai}
Here we report some additional details from our use of Fisherian randomization inference in the reanalysis of \citet{cai2015social}. In particular, Figure \ref{fig:cai_perm} shows the distribution of the Studentized H\`ajek estimator for the effect of seeding strategy on insurance takeup.

\begin{figure}[tb]
\centering
\includegraphics[scale=.8]{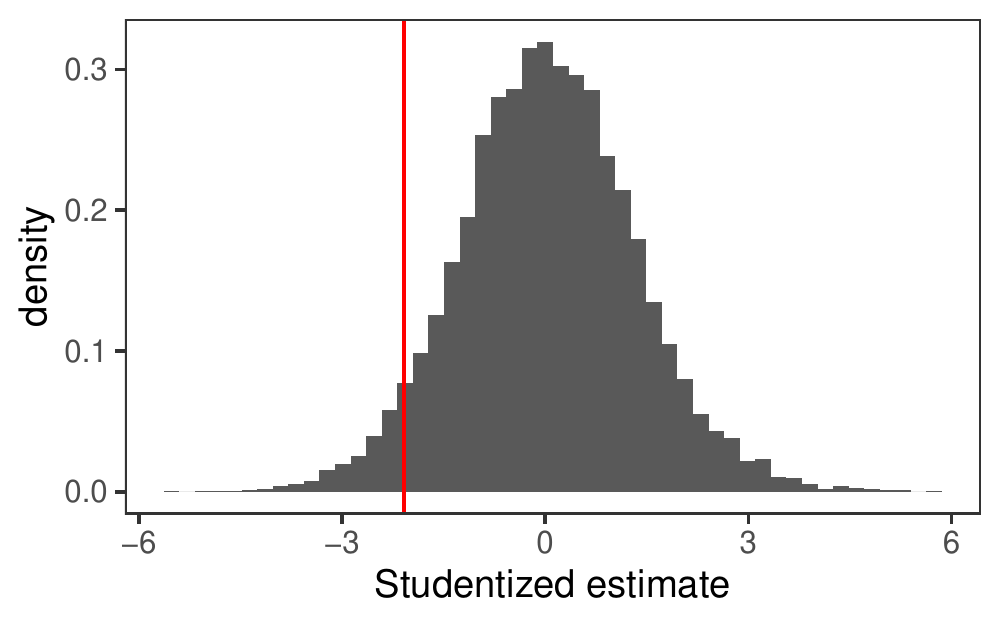} 
\caption{Fisherian randomization inference for insurance takeup using the distribution of Studentized H\`ajek estimator under a sharp null. This figure elaborates on Table \ref{table:cai-results}.  Observed statistic shown by red line.
}
\label{fig:cai_perm}
\end{figure}

\subsection{Paluck et al.~(2016)}
\label{app:paluck}
Here we report some additional details from the reanalysis of \citet{paluck2016changing}.

First, Figure \ref{fig:cai_perm} shows the distribution of the Studentized H\`ajek estimator for the effect of seeding strategy on insurance takeup.

\begin{figure}[h]
\centering
\includegraphics[scale=.8]{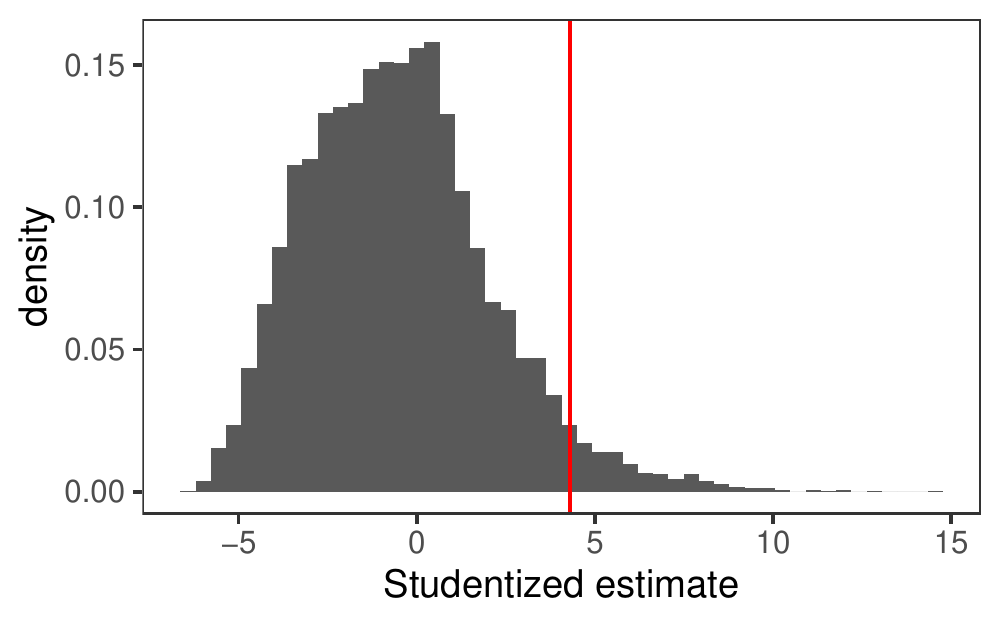} 
\caption{Fisherian randomization inference for peer conflict per student using the distribution of Studentized  under a sharp null. This figure elaborates on Table \ref{table:paluck-results}.  Observed statistic shown by red line.
}
\label{fig:paluck_perm}
\end{figure}

Here we provide results for two additional outcomes: the self-reported rate of wearing an anticonflict wristband and the self-reported number of friends talking about peer conflict.  An increase in both of these outcomes is viewed as a desirable result.

\begin{table}[H]
\centering
\begin{tabular}{rr}
  \hline
estimate (one-hop $-$ rand) & 0.0544 \\ 
  SE (analytic) & 0.0141 \\ 
  SE (bootstrap) & 0.0260 \\ 
  95\% CI (analytic) & [0.0269, 0.0820] \\ 
  95\% CI (bootstrap) & [-0.0075, 0.0886] \\ 
  p-value (analytic) & 0.00011 \\ 
  p-value (Fisherian) & 0.1376 \\ 
   \hline
\end{tabular}
\caption{H\'ajek estimate and inference for the difference in self-reported wristband-wearing one-hop and random targeting for \citet{paluck2016changing}.}
\label{table:paluck-results2}
\end{table}

\begin{table}[H]
\centering
\begin{tabular}{rr}
  \hline
estimate (one-hop $-$ rand) & 0.0600 \\ 
  SE (analytic) & 0.0190 \\ 
  SE (bootstrap) & 0.0256 \\ 
  95\% CI (analytic) & [0.0228, 0.0973] \\ 
  95\% CI (bootstrap) & [0.0111, 0.1153] \\ 
  p-value (analytic) & 0.0016 \\ 
  p-value (Fisherian) & 0.2706 \\ 
   \hline
\end{tabular}

\caption{H\'ajek estimate and inference for the difference in self-reported friends talking about peer conflict one-hop and random targeting for \citet{paluck2016changing}.}
\label{table:paluck-results3}
\end{table}

\begin{figure}[h]
\centering
A
\includegraphics[scale=.35]{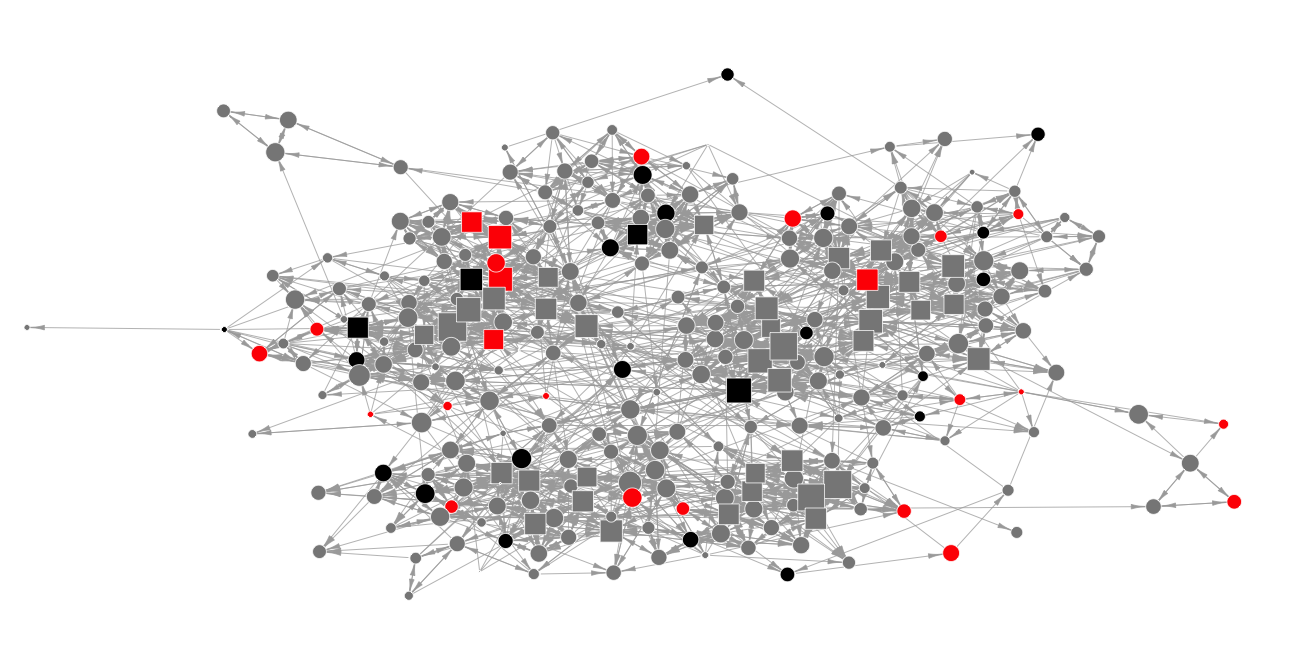} \\
B
\includegraphics[scale=.35]{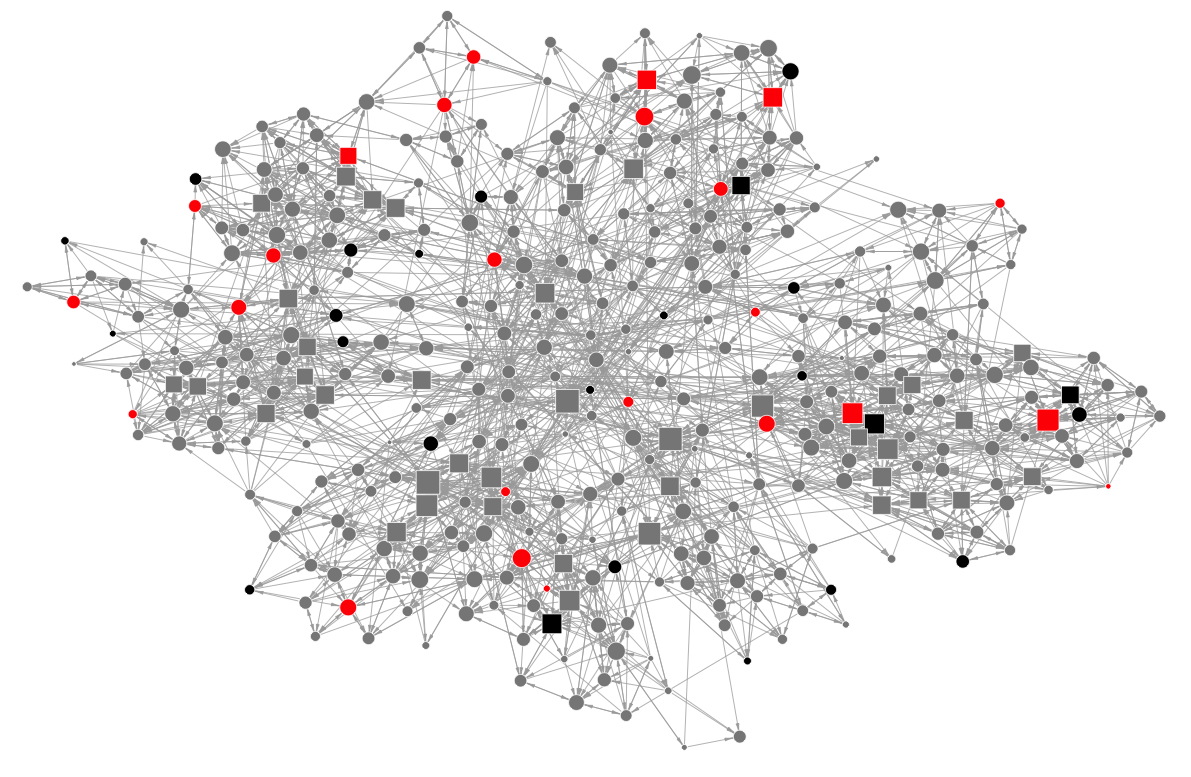}
\caption{Social networks for two schools in \citet{paluck2016changing} showing social referents (squares) and students eligible to be selected (black) and students selected as seeds (red).
Each node $v$ is sized proportional to $\P_i^{A,\text{repl}}(S_i = {v})$ (i.e., row-normalized in-degree), not accounting for being eligible for treatment.
Both have a somewhat similar fraction of the seed set who are social referents, with A a bit larger than B (A: 0.208, B: 0.167). But this is notably reversed for $w_A$ (A: 1.1e-4, B: 0.00395).
}
\label{fig:paluck_school_network_comparison}
\end{figure}
}

\end{document}